\documentclass[12pt,preprint]{aastex}
\RequirePackage{lineno}
\usepackage{amssymb}
\usepackage{amsmath}
\usepackage{mathrsfs}
\usepackage{natbib}
\usepackage[colorlinks, linkcolor=blue, urlcolor=blue, citecolor=blue]{hyperref}
\usepackage{array}
\usepackage{float}
\usepackage{graphicx}
\usepackage{subfigure}
\usepackage{rotating}
\usepackage{color}
\usepackage[all]{hypcap}
\usepackage{multirow}
\usepackage[toc,page]{appendix}

\begin{document}

\title{Constraining the long-lived magnetar remnants in short gamma-ray bursts from late-time radio observations}
\author{Liang-Duan Liu\altaffilmark{1}, He Gao\altaffilmark{1,*} and Bing Zhang\altaffilmark{2}}

\altaffiltext{1}{Department of Astronomy, Beijing Normal University, Beijing 100875, China; gaohe@bnu.edu.cn.}
\altaffiltext{2}{Department of Physics and Astronomy, University of Nevada, Las Vegas, NV 89154, USA.}

\begin{abstract}
	The joint detection of GW 170817 and GRB 170817A indicated that at least a fraction of short gamma ray bursts (SGRBs) originate from binary neutron star (BNS) mergers. One possible remnant of a BNS merger is a rapidly rotating, strongly magnetized neutron star, which has been discussed as one possible central engine for GRBs.  For a rapidly rotating magnetar central engine, the deposition of the rotation energy into the  ejecta launched from the merger could lead to  bright radio emission. The brightness of radio emission years after a SGRB would provide an  estimate of the kinetic energy of ejecta and, hence, a possible constraint on the BNS merger product. We perform a more detailed calculation on the brightness of radio emission from the interaction between the merger ejecta and circumburst medium in the magnetar scenario, invoking several important physical processes such as generic hydrodynamics, relativistic effects, and the deep Newtonian phase.
	 We use the model to constrain the allowed parameter space for 15 SGRBs that have late radio observations. Our results show that an injection energy of $E_{\rm inj} \sim 10^{52}$ erg is allowed for all the cases, which suggests that the possibility of a supra-massive or hyper-massive neutron star remnant is not disfavored by the available radio data. 
\end{abstract}


\section{Introduction}
The most promising model for short gamma-ray bursts (SGRBs) is mergers of two compact objects, such as double neutron stars (NS-NS) or a neutron star - black hole  (NS-BH) systems. The detection of the gravitational wave event GW 170817 from an NS-NS merger \citep{Abbott2017a}, and its associated short gamma-ray burst GRB 170817A \citep{Abbott2017b,Goldstein2017,Zhang2018} unambiguously confirmed that at least a fraction of SGRBs originate from binary neutron star mergers. However, whether or not a long-lived NS remnant could be formed during this merger event remains an open question \citep[e.g.][]{Ai2018,Ai2019}.

The recent discovery of a millisecond pulsar MSP J0740+6620,  with a mass  $2.14^{+0.10}_{-0.09}M_\odot$ \citep{Cromartie2019}, posed a strong constraint on the equation of state of high-density matter. This mass could be used as a lower limit on the maximum NS mass and rule out  soft equations of state (EOS) of NS that cannot produce such a high mass NS. For a relatively small total mass of a binary neutron star (BNS) system, a long-lived remnant could be formed \citep{Dai2006,Zhang2013,Giacomazzo2013}. A rapidly spinning magnetar has been suggested as the central engine of GRBs \citep{Duncan1992,Usov1992,Dai1998,Zhang2001,Dai2006}.  In the case of SGRBs, a long-lived magnetar can help to interpret several interesting X-ray activities following the GRBs, such as X-ray plateaus \citep{Dai1998,Zhang2001,Rowlinson2013},  extended emission \citep{Metzger2008}, and X-ray flares \citep{Dai2006}.  \cite{Ciolfi2019} recently performed general relativistic magnetohydrodynamic (GRMHD) simulations of a BNS merger system  up to $\sim 100$ ms after the merge and followed the evolution of the rotational and magnetic energy of a long-lived magnetar in great detail. 

Numerical simulations of NS-NS mergers indicated the typical masses of the merger ejecta  $M_{\rm ej} \sim 10^{-3}M_{\odot}$ to a few $10^{-2}M_{\odot}$ , and the velocities of the ejecta  $v_{\rm ej} \sim 0.1-0.3c$ \citep[e.g.][]{Rezzolla2010,Rosswog2013,Hotokezaka2013,Siegel2017}. The interaction between the sub-relativistic merger ejecta with the surrounding medium would give rise to synchrotron radio emission on longer timescales $\sim$ a few years \citep{Nakar2011,Gao2013,Hotokezaka2015}.  If the merger remnant is a rapidly rotating magnetar,  richer electromagnetic signals are expected. These include GRB-less X-ray transients \citep{Zhang2013,Sun2017}, magnetar-boosted kilonova-like events known as ``merger-novae'' \citep{Yu2013,MetzgerPiro2014,gao15a,gao17}, and the brighter forward and reverse shock emission from the interaction  between the engine-powered ejecta and the surrounding medium \citep{Gao2013,Wang2015,Liu2016}. Recently, a GRB-less X-ray transient CDF-S XT2
was reported by \cite{Xue2019}, which can be interpreted as originating from the internal magnetic
dissipation process in an ultra-relativistic wind of a newborn magnetar \citep{Xiao2019,Sun2019}.

A magnetar would deposit a significant fraction of its rotational energy into the merger ejecta to increase its kinetic energy. 
Radio observations on the timescale of $\sim$ years after the bursts provide a probe of the total kinetic energy of ejecta.   Several groups tried to search for late-time radio emission following SGRBs and use the non-detection upper limits to constrain the existence of a magnetar central engine \citep{Metzger2014,Horesh2016,Fong2016,Klose2019}. 
An upper limit of a few times $10^{51}$ erg of kinetic energy was claimed for some SGRBs, which was used to argue against a magnetar engine \citep{Horesh2016,Fong2016}. However, there is a high level of degeneracy between the kinetic energy and other model parameters. For example, in these calculations, large values of the shock microscopical parameters (e.g., $\epsilon_B = 0.1$) have been adopted. In addition, some simplifications of the model have been adopted (e.g. in \cite{Metzger2014} and \cite{Fong2016}), which led to tighter constraints on the magnetar model.

In order to more precisely calculate the radio emission flux following SGRBs in the timescale of $\sim 1-10$ years after the bursts,  we developed a more sophisticated model by invoking several important physical processes not fully incorporated in previous models, e.g, generic hydrodynamics, relativistic effects, and the deep Newtonian phase. We collect the late-time radio observational data of 15 SGRBs from the literature and constrain the allowed parameter space for a long-lasting NS as the BNS merger remnant using the observations. In Section \ref{Sec:Model}, we describe our model in detail. In Section \ref{Sec:Application}, we show the applications of our model to the observations. Our conclusions and discussion are presented in Section \ref{Sec: Discussion}.


\section{Model} 
\label{Sec:Model}
If the equation of state (EOS) of neutron stars (NSs) is stiff enough, at least a fraction of the BNS mergers will leave behind a supra-massive or even a stable NS that spins rapidly with a strong magnetic field \citep{Dai2006,Zhang2013,Gao2016,Piro2017,Margalit2019}. Such a magnetar would deposit a significant fraction of its rotational energy into the merger ejecta. The kinetic energy of the merger ejecta would significantly increase. The interaction between the merger ejecta and the ambient medium produces radio emission via synchrotron radiation of relativistic electrons. Due to the additional energy injection from the long-lasting magnetar remnant, the radio brightness would be significantly enhanced \citep{Gao2013,Metzger2014,Horesh2016,Fong2016}. 


The rotational energy of an NS formed by a BNS merger is 
\begin{equation}
E_{\rm rot}=\frac{1}{2}I \Omega^2 \simeq 2 \times 10^{52}  I_{45}\left( \frac{P_0}{1 \rm{ ms}} \right)^{-2} \rm{ergs},
\label{eq:Erot}
\end{equation}
where $I$ is the moment of inertia of the proto-NS, and for a massive NS formed from a BNS merger, one has $I_{45} \sim 1.5$. All quantities are in c.g.s units and the convention $Q_n = Q/10^n$ has been adopted throughout the paper. Because the merging BNS has a high orbital angular momentum, the post-merger proto-NS would be rotating extremely rapidly, with an initial rotation period close to the centrifugal breakup limit, e.g. $P_0 \sim 1$ ms. The rotation energy $E_{\rm rot}$ in Eq.(\ref{eq:Erot}) presents a characteristic energy for the magnetar model to be tested with the radio data. Since the millisecond pulsar energy injection is essentially isotropic, the injected energy can be regarded as the isotropic equivalent energy in the ejecta - medium interaction model discussed in the rest of the paper.

Due to the dissipation of the newborn magnetar wind, a fraction of $E_{\rm rot}$ would be radiated to power early bright X-ray and optical emissions \citep{Zhang2013,Sun2017}. It is possible that some fractions of the energy is radiated by secular gravitational waves \citep{Fan2013,Gao2016} or fall into the black hole for a supramassive NS that collapses before fully spinning down \citep{Gao2016}. In any case, a good fraction of the rotation energy would be transferred into the merger ejecta, as $E_{\rm inj} = \xi E_{\rm rot}$, where $\xi<1$ is the fraction of rotation energy that is injected into the shock. Whether or not the ejecta can be accelerated to a relativistic speed depends on $E_{\rm inj}$ and the ejecta mass $M_{\rm ej}$. With $ E_{\rm inj} \sim M_{\rm ej} c^2$, one can define a characteristic ejecta mass \citep{Gao2013}
\begin{equation}
M_{\rm{ej,c}} \sim 1.1 \times 10^{-2}M_{\odot}\xi \left( \frac{E_{\rm {rot}}}{2 \times 10^{52} \rm{\ ergs}} \right).
\end{equation}
An ejecta lighter than $M_{\rm{ej,c}}$ can be accelerated to a relativistic speed. For such a case, some relativistic effects should  be taken into account. 

In order to calculate the blast wave dynamics in both relativistic and non-relativistic (Newtonian) phases, we use the generic dynamical model proposed by \cite{Huang1999}\footnote{More precise generic dynamical models have been later proposed with increasing sophistication \citep[e.g.][]{peer2012,Nava2013,Zhang2018-book}. However, for the purpose of this work, the simpler model of \cite{Huang1999} suffices.}.  Consider the energy injection from the magnetar and deceleration of the ejecta due to interaction with circumburst medium. The bulk Lorentz factor of the shock $\Gamma$ evolves with the ejecta radius $R$ as \citep{Liu2014}
\begin{equation} \label{eq:dy1}
  \frac{d \Gamma}{d R} = \frac{4 \pi R^2 n m_p}{M_{\rm{ej}} + 2 \Gamma
  M_{\rm{sw}}} \left[ \frac{L_{\rm{inj}} (t)}{c^2} \frac{d t}{d
  M_{\rm{sw}}} - (\Gamma^2 - 1) \right],
\end{equation}
where $n$ is the number density of the surrounding medium, $m_p$ is the proton mass, $c$ is the speed of light, and $L_{\rm{inj}} (t) $ is the injected luminosity from the magnetar. We characterize the injection luminosity as $L_{\rm{inj}} (t) = \xi L_{\rm{sd}} (t)$. Assuming that the main channel of proto-magnetar energy loss is via dipole radiation, the spin-down luminosity can be written as $L_{\rm sd}=L_{\rm sd,0}(1+t/T_{\rm sd})^{-2}$. The characteristic spin-down luminosity $L_{\rm sd,0}$ and time scale $T_{\rm sd}$ critically depend on the magnetic field strength of the magnetar (given a particular $E_{\rm rot}$ which is defined by the initial period $P_0$). 

The evolution of the mass of the  swept-up medium $M_{\rm{sw}}$ and the radius of
the ejecta $R$ are given by \citep{Huang1999}
\begin{equation} \label{eq:dy2}
  \frac{d M_{\rm{sw}}}{d R} = 4 \pi R^2 n m_p,
\end{equation}
and
\begin{equation} \label{eq:dy3}
  \frac{d R}{d t} = \frac{\beta c}{1 - \beta},
\end{equation}
where $\beta$ is the velocity of the ejecta divided by the speed of light $c$. Initially, the kinetic energy of the ejecta would increase because of energy injection. When the ejecta collects a mass comparable to its own, the shock begins deceleration at the characteristic timescale 
\begin{equation}
t_{\rm {dec}} \sim 4.9 \times 10^{5}   \xi^{-7/3} E_{\rm {rot, 52}}^{-7/3} M_{ \rm {ej,-3}}^{8/3} n^{1/3} \rm{s}.
\end{equation}
The radio lightcurve usually peaks at the timescale 
$t \sim t_{\rm dec}$ \citep{Nakar2011,Metzger2014}, so the radio observations at this timescale offer an important probe of the total kinetic energy in the shock.

\begin{figure}[tbph]
\begin{center}
\includegraphics[width=0.48\textwidth,angle=0]{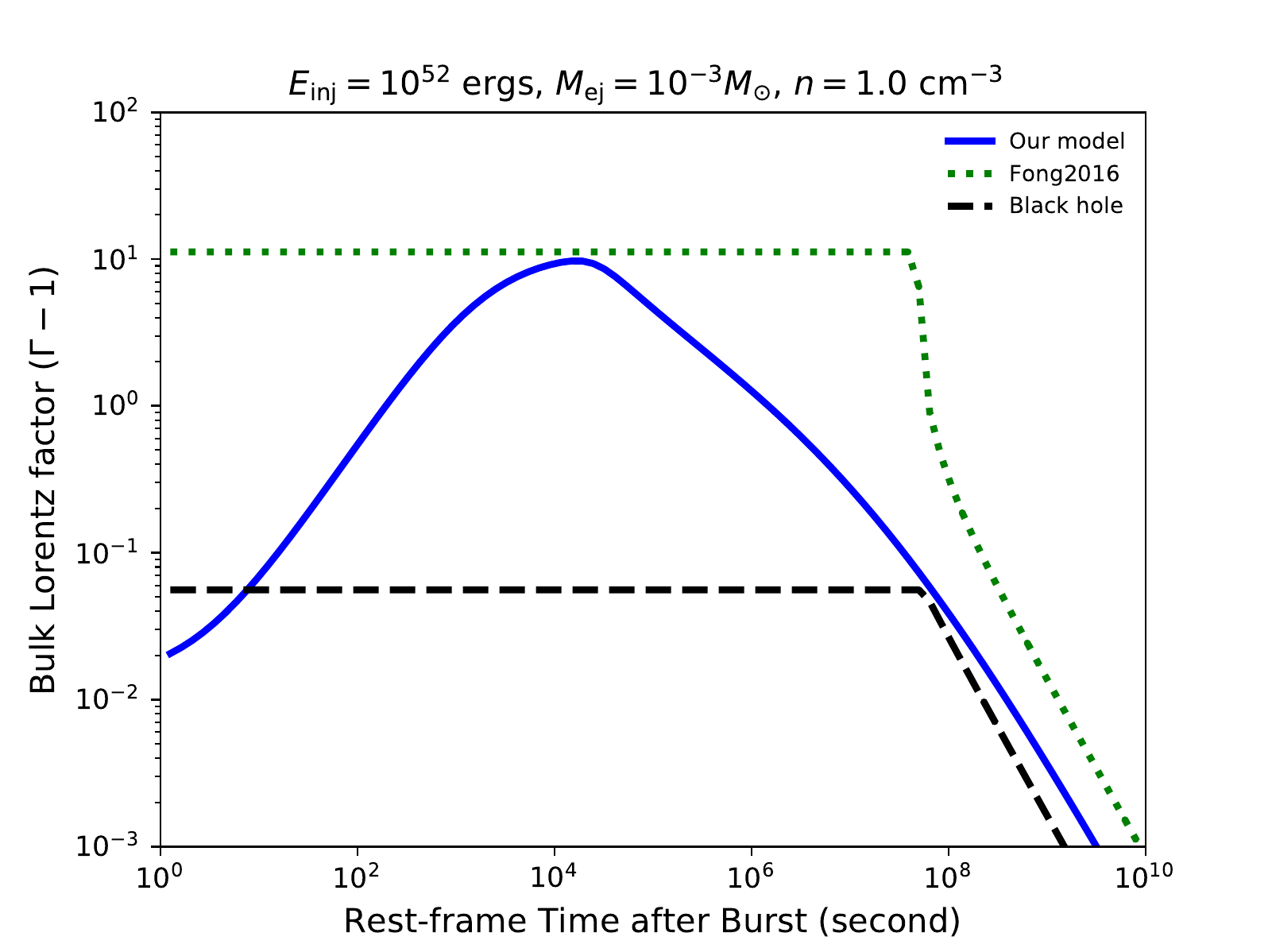}
\includegraphics[width=0.48\textwidth,angle=0]{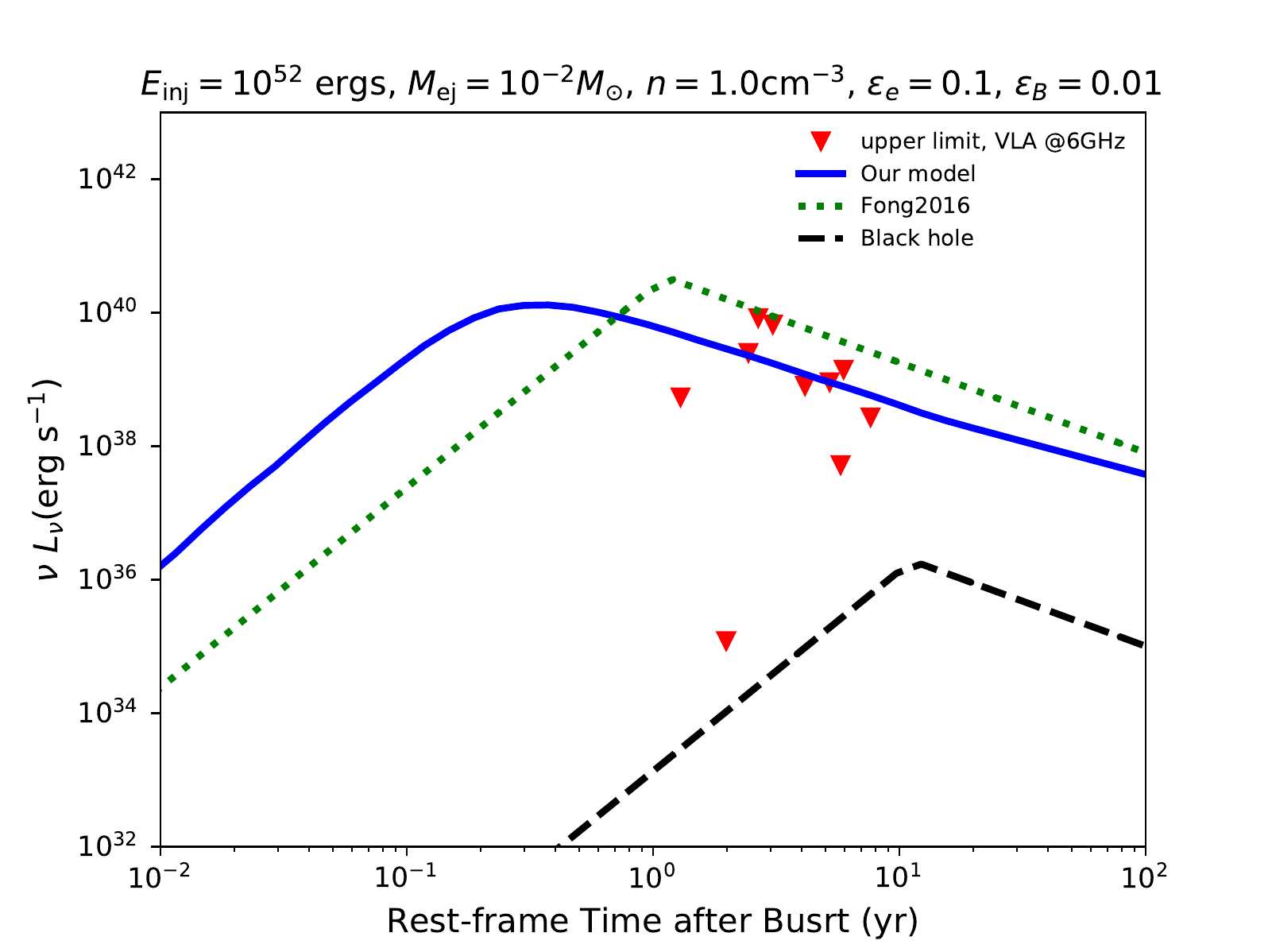}
\end{center}
\caption{A comparison of the numerical results of our model with \cite{Fong2016} and \cite{Nakar2011}. The left panel is the evolution the bulk Lorentz factor of the ejecta, the right panel is the radio light curves at 6 GHz. Same values for physical parameters have been adopted to ensure a uniform comparison. The black hole model has a total kinetic energy $E_k = 10^{50}$ erg.}
\label{fig:dynamics}
\end{figure}

\cite{Nakar2011} calculated synchrotron radio emission lightcurve 
in the black hole scenario without energy injection from the central engine. They adopted the kinetic energy of the ejecta $E_{\rm k} \sim 10^{49}-10^{50}$ erg. The velocity of the ejecta is non-relativistic. The shock is in
free coasting phase early on and enters the subsequent Sedov-Taylor self-similar evolution later. \cite{Metzger2014} and \cite{Fong2016} used the  similar method to calculate the dynamical evolution for the case of a magnetar. They used the rotation energy of the magnetar $E_{\rm rot}$ instead of the kinetic energy of the ejecta $E_{\rm k}$ in the calculations.
Using Equations (\ref{eq:dy1})-(\ref{eq:dy3}), we can calculate the evolution of the bulk Lorentz factor of the ejecta with the generic hydrodynamics model. The numerical results are shown in the left panel  of Fig. \ref{fig:dynamics}. The blue solid line is the evolution of the ejecta bulk Lorentz factor based on the generic hydrodynamics model, and the green dotted line is the dynamical evolution adopted by \cite{Fong2016}. We find that the two models obtain the same maximum bulk Lorentz factor of the ejecta $\Gamma_{\max} \approx \xi E_{\rm{rot}}/M_{\rm{ej}}c^2$, but the deceleration time-scale in our model is shorter than that of \cite{Fong2016}.

In the synchrotron blast-wave model \citep{Sari1998}, the observed spectra reflect that the distribution of the shock-accelerated electrons Lorentz factor $\gamma_e$. It is usually assumed that the electron energy spectrum is a power law with slope $p$, i.e.
\begin{equation}
d N \propto \gamma_e^{- p} d \gamma_e, \qquad \gamma_e \geqslant \gamma_m
\end{equation}
for a mildly relativistic shock $p \approx 2.1 - 2.5$ \citep{Nakar2011}.
 The minimum electron Lorentz factor can be obtained based
on the total energy of the accelerated electron, i.e.
\begin{equation}
\gamma_m - 1= \frac{p - 2}{p - 1} \frac{m_p}{m_e} \epsilon_e (\Gamma - 1),
\end{equation}
where $\epsilon_e$ is the fraction of the total internal energy of the shocked
medium carried by electrons.  

It is common to relate the magnetic field energy density ($B^2/8\pi$) and the internal energy of post-shocked medium ($U'$) with a shock microphysics parameter  $\epsilon_B$. Based on the relativistic shock jump conditions, the internal energy density of the shocked medium  can be written in the form of $U'=  (4 \Gamma + 3) (\Gamma - 1) n m_p c^2$. Under this assumption, the  
magnetic field strength in the shock can be estimated by \citep{Sari1998}
\begin{equation}
B = \sqrt{8 \pi \epsilon_B (4 \Gamma + 3) (\Gamma - 1) n m_p c^2}.
\end{equation}

The late time radio spectrum produced by the shock is irrelevant to the
cooling frequency $\nu_c$. The spectrum is determined by two characteristic
frequencies, one is the typical synchrotron of electrons $\nu_m$ with the minimum electron
Lorentz factor $\gamma_m$, i.e.
\begin{equation}
\nu_m \simeq \frac{3}{4 \pi} \Gamma \gamma_m^2 \frac{e B}{m_e c},
\end{equation}
The factor of $\Gamma$ is introduced to transfer the shock co-moving frame to the frame of the observer.
The other one is the synchrotron self-absorption frequency $\nu_a$, which can be
estimated by requiring that the optical depth equal to unity. In the case we are
interested in,  $\nu_a$ can be expressed as \citep{Zhang2018-book}
\begin{equation}
\nu_a = \left\{ \begin{array}{ll}
\left( \frac{\mathcal{C}_1 e n R}{B \gamma_m^5} \right)^{3 / 5} \nu_m, &
\nu_a < \nu_m,\\
\left( \frac{\mathcal{C}_2 e n R}{B \gamma_m^5} \right)^{2 / (p + 4)}
\nu_m, & \nu_a \geqslant \nu_m .
\end{array} \right.
\end{equation}
where the coefficients depend on the electron power law index $p$, i.e.
\begin{eqnarray}
\mathcal{C}_1 = \frac{16 \pi \sqrt[3]{4}}{3 \Gamma \left( \frac{1}{3}
	\right)} \frac{p + 2}{3 p + 2}, & \rm{and} & \mathcal{C}_2 = \frac{2^{3 +
		\frac{p}{2}}}{3 \sqrt{3}} \Gamma \left( \frac{1}{6} + \frac{p}{4} \right)
\Gamma \left( \frac{11}{6} + \frac{p}{4} \right).
\end{eqnarray}
The peak specific synchrotron emission power of a single electron in the
observer frame can be expressed as \citep{Sari1998}
\begin{equation}
P_{\nu, \max} = \frac{m_e c^2 \sigma_T}{3 e} \Gamma B,
\end{equation}
which is independent of the electron Lorentz factor $\gamma_e$.  The total
number of the swept-up electrons in the post-shock gas is $N_e = 4 \pi R^3 n / 3.$
The observed peak flux at the luminosity distance $D_L$ can be written as
\begin{equation}
F_{\nu, \max} =(1+z) \frac{N_e P_{\nu, \max}}{4 \pi D_L^2} .
\end{equation}
The radio-band synchrotron spectrum from the shock is governed by the relative orderings between $\nu_a $ and $\nu_m$.  There are two possible types of the radio spectra, see \cite{Gao2013b} and \cite{Piran2013} (their Figure 4):
for $\nu_a < \nu_m$, the observed flux at an observational frequency $\nu_{\rm{obs}}$ is given by
\begin{equation}
F_{\nu} = F_{\nu, \max} \left\{ \begin{array}{ll}
\left( \frac{\nu_a}{\nu_m} \right)^{\frac{1}{3}} \left(
\frac{\nu_{\rm{obs}}}{\nu_a} \right)^2, & \nu_{\rm{obs}} \leqslant
\nu_a < \nu_m,\\
\left( \frac{\nu_{\rm{obs}}}{\nu_m} \right)^{\frac{1}{3}}, & \nu_a <
\nu_{\rm{obs}} \leqslant \nu_m,\\
\left( \frac{\nu_{\rm{obs}}}{\nu_m} \right)^{- \frac{p - 1}{2}}, & \nu_a
< \nu_m < \nu_{\rm{obs},}
\end{array} \right.
\end{equation}
and for $\nu_a > \nu_m$, the observed flux $F_{\nu}$ is
\begin{equation}
F_{\nu} = F_{\nu, \max} \left\{ \begin{array}{ll}
\left( \frac{\nu_m}{\nu_a} \right)^{\frac{(p + 4)}{2}} \left(
\frac{\nu_{\rm{obs}}}{\nu_m} \right)^2, & \nu_{\rm{obs}} \leqslant
\nu_m < \nu_a,\\
\left( \frac{\nu_a}{\nu_m} \right)^{- \frac{(p - 1)}{2}} \left(
\frac{\nu_{\rm{obs}}}{\nu_a} \right)^{\frac{5}{2}}, & \nu_m <
\nu_{\rm{obs}} \leqslant \nu_a,\\
\left( \frac{\nu_{\rm{obs}}}{\nu_m} \right)^{- \frac{p - 1}{2}}, & \nu_m
< \nu_a < \nu_{\rm{obs}.}
\end{array} \right.
\end{equation}


As the shock wave sweeps across the ambient circumburst medium, the shock slows down to a
non-relativistic speed (i.e., $\Gamma -1 \ll 1$). The dynamics can be then described by the non-relativistic Sedov-Taylor self-similar solution, $\beta \propto t^{-3/5} $. If the minimum electron Lorentz factor still satisfies $\gamma_m \gg 1$, the synchrotron flux in the radio band would decay as  $F_{\nu} \propto t^{- 3 (5 p - 7) / 10}$ \citep{Frai2000}.
Once the majority of the shock accelerated electrons are no longer highly relativistic, the blast wave would enter the
so called ``deep Newtonian phase'' as studied by \cite{Huang2003}.  In this situation, according to the theory of Fermi acceleration in non-relativistic shock, the electron spectrum is likely to be a power-law distribution in the momentum space rather than in the energy space \citep{Sironi2013}.

The deep Newtonian phase would begin at the time $t_{\rm DN}$ when $\gamma_m -1 \sim 1$ \citep{Sironi2013}, corresponding to the velocity of the shock $\beta \sim 0.22 \epsilon^{- 1 / 2}_{e, - 1}$. This is at
\begin{equation}
t_{\rm DN} \sim 370  \xi E_{\rm 52}^{1/3} n^{-1/3} \rm{\ days}.
\end{equation}
When $t>t_{\rm DN}$, most of the electron energy is contributed by the electrons with $\gamma_e \sim 2$ and the electron spectrum follows a power law distribution in the momentum space.  In the deep Newtonian regime, the radio flux decay as $F_{\nu} \propto t^{-3(1+p)/10}$ \citep{Granot2006,Sironi2013}.  This temporal index is shallower than the one derived by ignoring this effect.

The comparisons of our model with the four previous relevant works (i.e., \cite{Nakar2011}, \cite{Metzger2014}, \cite{Fong2016}, and \cite{Horesh2016}) are given in Table. \ref{tb:models}. In these previous papers, some of the important physical processes discussed here were not  taken into account.  

The numerical results are shown in Fig. \ref{fig:dynamics}. The radio lightcurve peak time at 6 GHz calculated by our model (the blue solid line) is about one order of magnitude earlier than that of \cite{Fong2016}.  On the timescale of $\sim 1-10$ years after the bursts, which are the time windows for the observations, the theoretical luminosity calculated by our model is about one order of magnitude lower than the lightcurves predicted in \cite{Fong2016}, and several orders of magnitude higher than the black hole case \citep{Nakar2011}. With the detailed treatment of the ``deep Newtonian'' phase, at late times  the decline rate predicted in our model is shallower than those presented in  \cite{Fong2016} and  \cite{Nakar2011}.

\section{Application to SGRBs}
\label{Sec:Application}

We collect 15  SGRBs with radio observations on timescales of $\sim$ years from the literature \citep{Metzger2014,Horesh2016,Fong2016,Hajela2019}.  No radio source was detected in either case, and upper limits of the radio flux $F_\nu$  at the level of $(8.4-510) \ \mu$Jy on the timescale of $189-3500$ days after the bursts were obtained, which correspond the luminosity upper limits $\nu L_\nu \sim (1.18 \times 10^{35}-1.83 \times10^{40}) \ {\rm erg \ s^{-1}}$ (Table \ref {tb:sample}).  
 
All the events in our sample have prior observations showing X-ray excess emission that could be a sign of the existence of a magnetar central engine (see column 7 of Table \ref {tb:sample}).  Nine events have extended emission, and seven show an X-ray plateau. The X-ray afterglow of GRB 100117A shows both an X-ray plateau and flares. In particular, \cite{Lv2015} fitted the X-ray lightcurves of GRB 050724A and GRB 090510 with an ``internal plateau'' model.  GRB 170817A displays an extended emission and a low-significance temporal feature in the X-ray afterglow, which is consistent with the reactivation of the central NS \citep{Piro2019}.
Three events (i.e., GRB 050724, GRB 051221A, and GRB 130603B)  have two radio observations on different frequencies and different times.

\begin{figure}[tbph]
\begin{center}
\includegraphics[width=0.95\textwidth,angle=0]{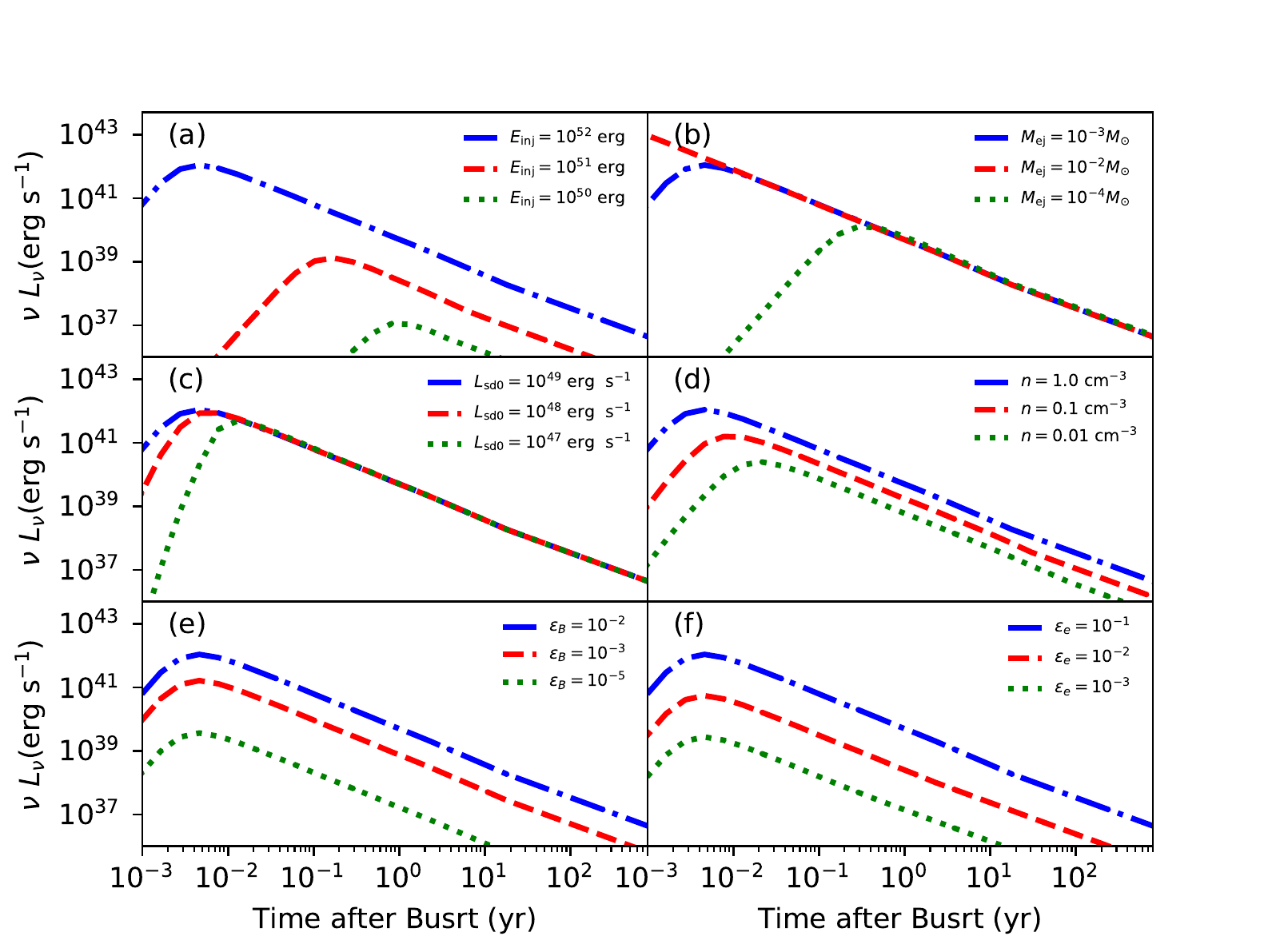}
\end{center}
\caption{Radio light curves at 6 GHz by varying various parameters:  injected energy from the magnetar $E_{\rm inj}$ (panel a),  mass of the merger ejecta $M_{\rm ej}$  (panel b),  initial spin-down luminosity $L_{\rm sd,0}$ (panel c), density of the surrounding medium $n$ (panel d), fraction of the post-shock energy density in magnetic field $\epsilon_B$ (panel e), and fraction of the post-shock energy density in electron $\epsilon_e$ (panel f).
 The fiducial parameters are (plotted with blue dash-dot  line): $E_{\rm inj}=10^{52}$ erg, $M_{\rm ej}=10^{-3}M_{\odot}$, $L_{\rm sd,0}=10^{48}$ erg s$^{-1}$, $n=1.0$ cm$^{-3}$, $\epsilon_B=10^{-2}$, $\epsilon_e = 0.1$, and $p=2.3$. }
\label{fig:parameters}
\end{figure}

The free parameters in our model include the injected energy from the magnetar $E_{\rm inj}$, the mass of the merger ejecta $M_{\rm ej}$, the initial spin-down luminosity $L_{\rm sd,0}$, the density of the surrounding medium $n$, the power law index of electron distribution $p$, and the 
shock microphysics parameters
$\epsilon_B$ and $\epsilon_e$. How various parameters might affect the properties of radio emission are shown in Fig. \ref{fig:parameters}. We find that there is a high level of degeneracy between the model parameters.

\begin{figure}[tbph]
\begin{center}
\includegraphics[width=0.47\textwidth,angle=0]{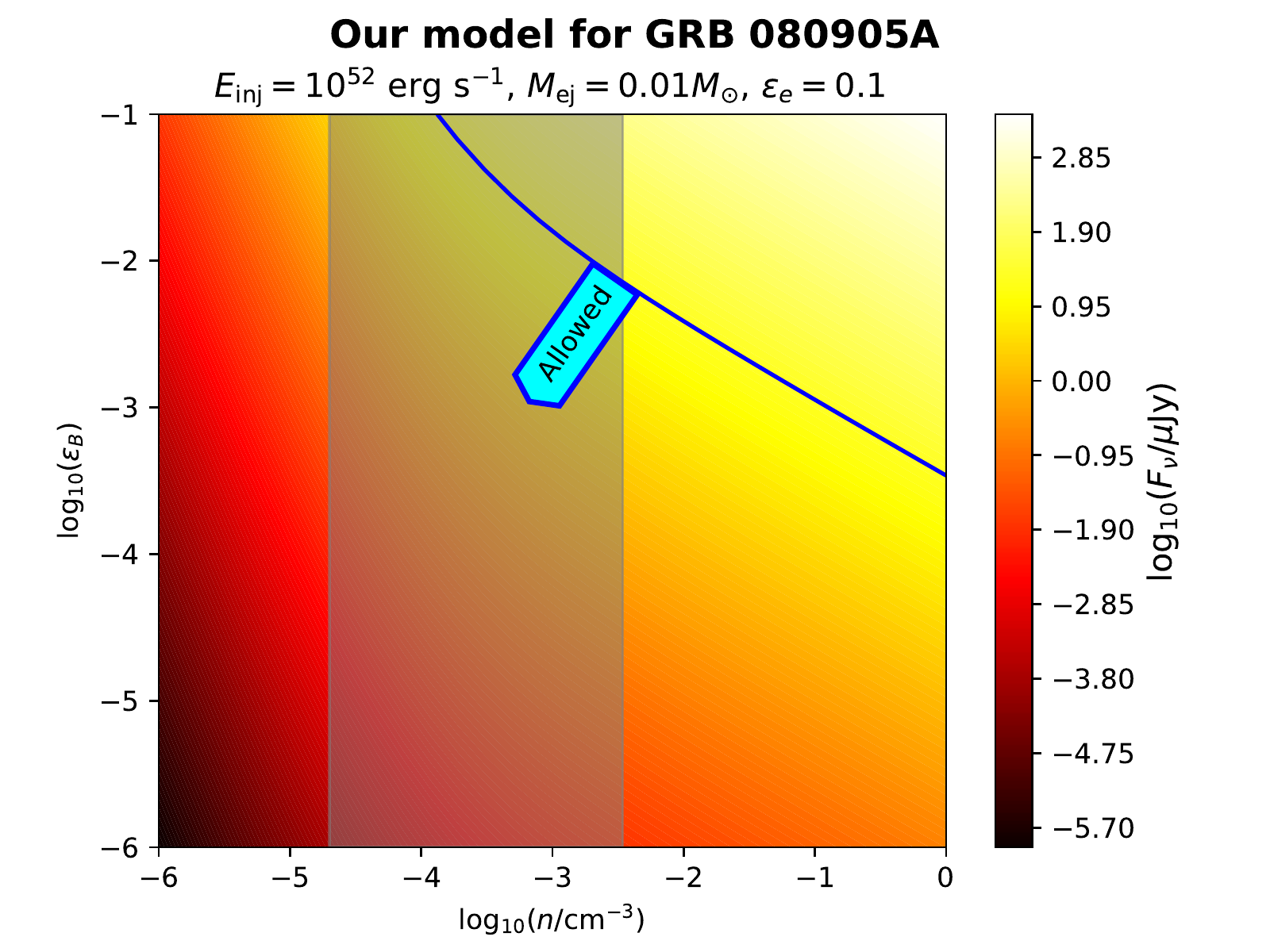}
\includegraphics[width=0.47\textwidth,angle=0]{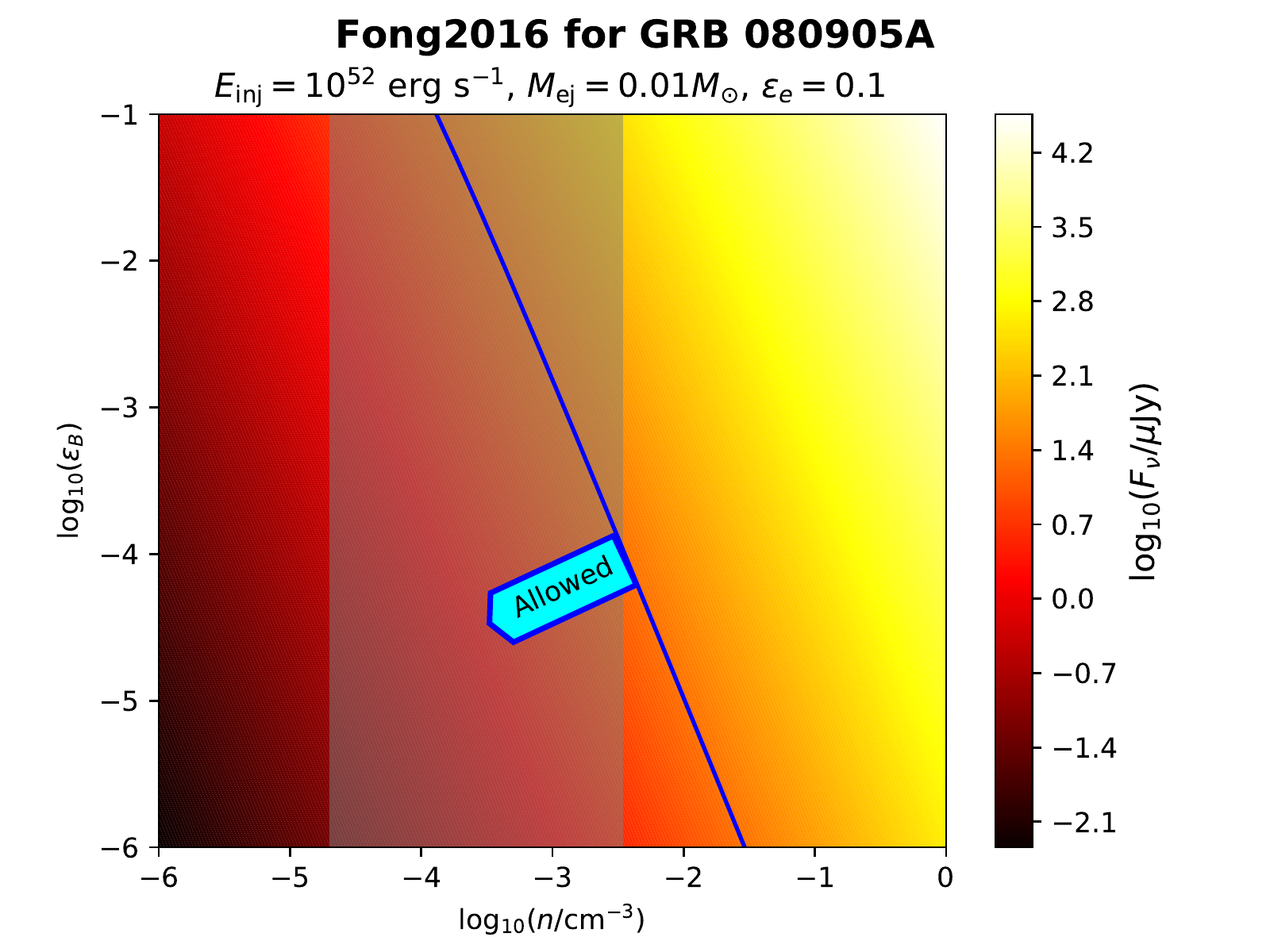}
\end{center}
\caption{The parameter space in the $n-\epsilon_B$ plane with color indicating the observed flux $F_\nu$. The lower-left part of each panel is the allowed parameter space from the non-detection of radio emission from GRB 080905A. The left panel is  the constraints based on our model and the right panel is the constraints based on \cite{Fong2016}. The light gray vertical region is the range of the allowed circumburst medium density independently determined from the broadband afterglow modeling. }
\label{fig:contour-GRB080905A}
\end{figure}

In Fig. \ref{fig:contour-GRB080905A}, we show the constraint on the long-lived magnetar from the upper limit of GRB 080905A.   Assuming the injected energy from the magnetar $E_{\rm inj} = 10^{52}$ erg, the ejecta mass $M_{\rm ej} =0.01M_{\odot} $ and $\epsilon_e =0.1$,  we present the parameter space in the $n-\epsilon_B$ plane with the color indicating the contours of the observed flux $F_\nu$. The lower-left part of each panel is the allowed parameter space from the non-detection of radio emission from GRB 080905A. Since the theoretical luminosity predicted in our model is one order of magnitude lower than that of  \cite{Fong2016} at the observational time  of this event, $T_{\rm rest}=5.769$ yr, our model predicts a larger allowed parameter space for the magnetar model to survive. 
 
\begin{figure}[tbph]
	\begin{center}
		\includegraphics[width=0.47\textwidth,angle=0]{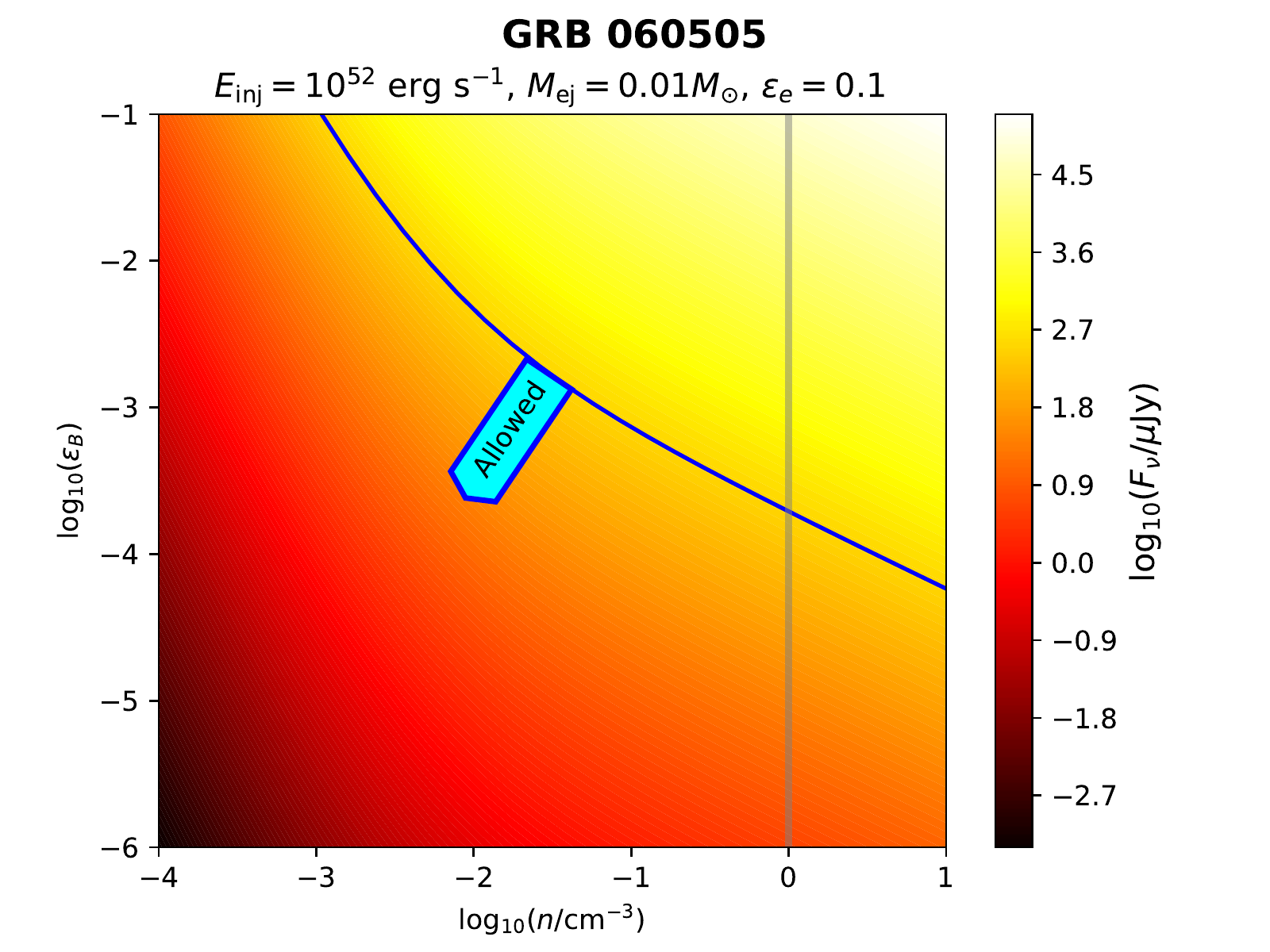}
		\includegraphics[width=0.47\textwidth,angle=0]{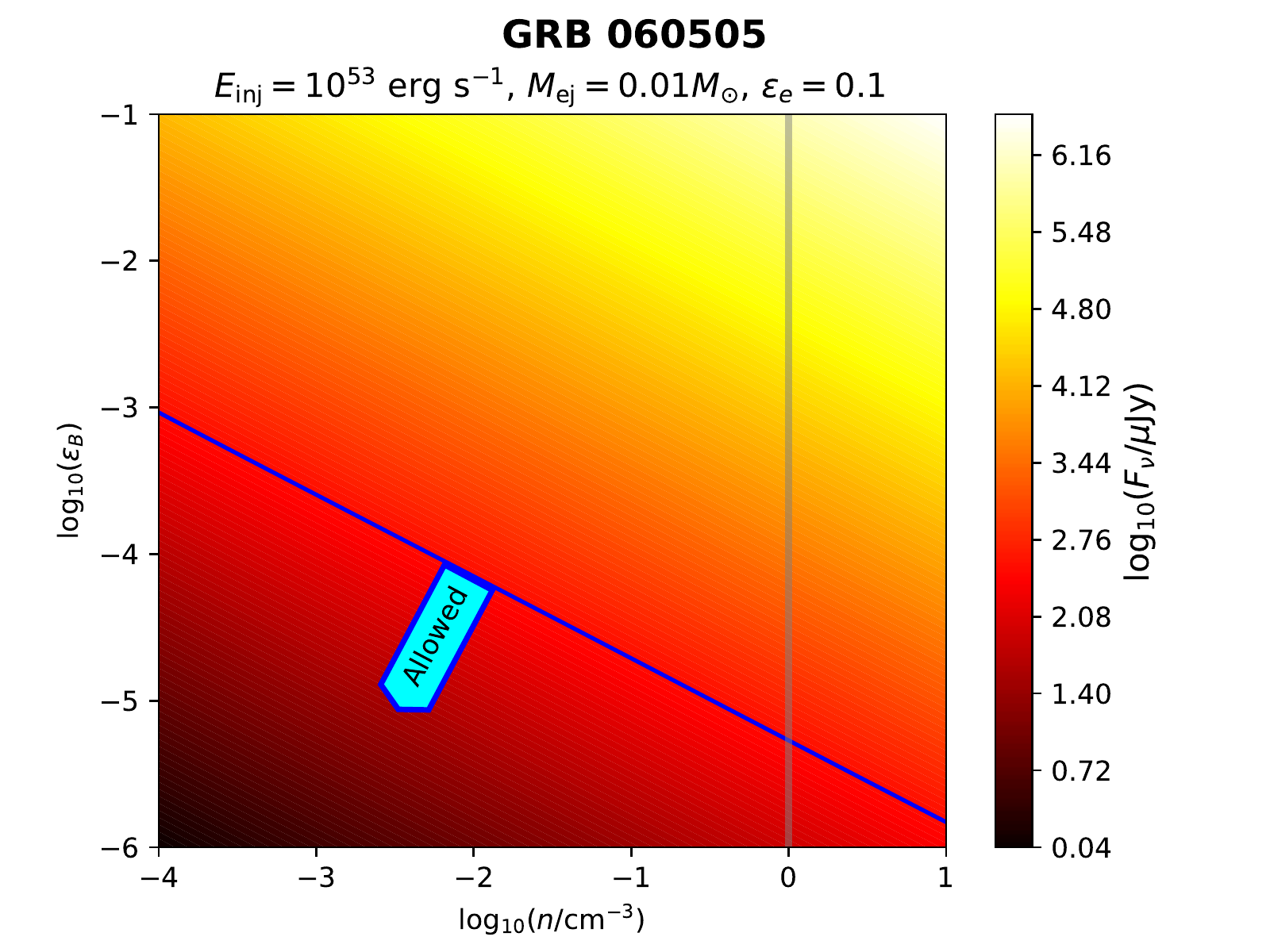}
		\includegraphics[width=0.47\textwidth,angle=0]{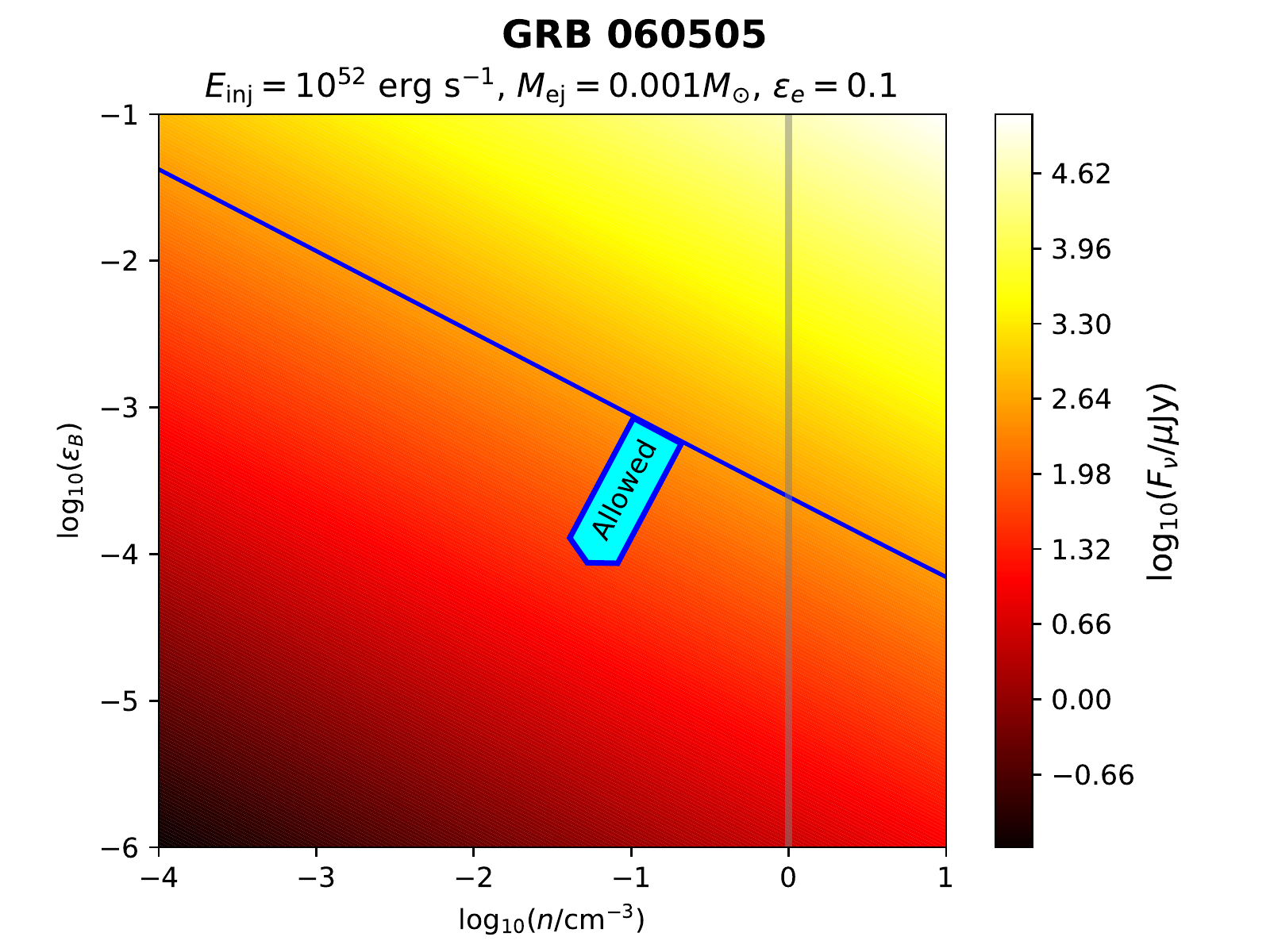}
		\includegraphics[width=0.47\textwidth,angle=0]{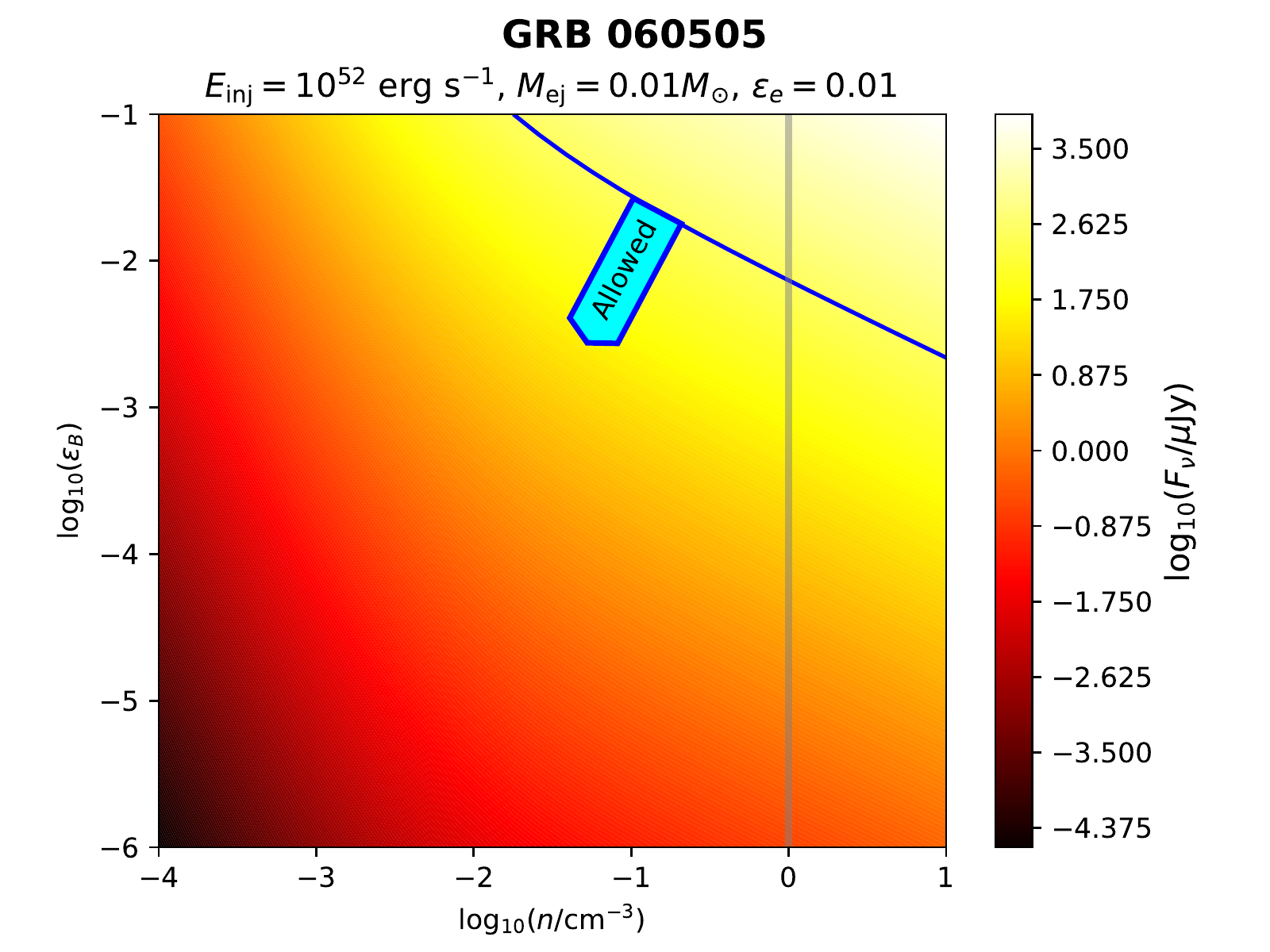}
	\end{center}
	\caption{Constraints on the parameter space with the non-detection radio emission from GRB 060505. The upper left panel is for fiducial parameters. The other three panels vary one parameter each to show the dependences of the parameters. }
		\label{fig:contour-GRB060505}
\end{figure}

Due to the high degeneracy between model parameters, adopting different values of $E_{\rm inj}$, $M_{\rm ej}$, and $\epsilon_e$ would change the allowable parameter space for the same observational upper limit. The constraints on the parameter space with the non-detection radio emission from GRB 060505 are shown in Fig. \ref{fig:contour-GRB060505}.  In the upper-left panel, we take  $E_{\rm inj} = 10^{52}$ erg, $M_{\rm ej} =0.01M_{\odot}$, and $\epsilon_e =0.1$ as the fiducial values of parameters. In each plot we vary one parameter while keeping the other parameters to the fiducial values. Afterglow modeling of GRB 060505 by \cite{Xu2009} suggested that the surrounding medium density is $n \sim 1$ cm$^{-3}$. From the constraint of the upper-left panel, one can estimate the maximum magnetic field fraction 
$\epsilon_{B, \max} \approx 3.7 \times 10^{-4}$ in the magnetar scenario that is allowed to satisfy the radio upper limit of GRB 060505. In the upper-right panel, we increase the injected energy by a factor of 10 to $E_{\rm inj} =10^{53}$ erg, reaching a tighter constraint on the allowed parameter space in the  $n-\epsilon_B$ plane. In the lower-left panel, we take a lower value of the ejecta mass $M_{\rm ej} =10^{-3}M_{\odot}$.
Compared with the fiducial parameters, this case has a slightly smaller allowed parameter space. In the lower-right panel, we adopt a lower value of $\epsilon_e =0.01$ \footnote{\cite{Gao2015} systematically investigated the \textit{Swift} GRB that have optical detections earlier than 500 s and found that the preferred electron equipartition parameter $\epsilon_e$ value is 0.01, which is smaller than the commonly used value.}. As shown in the panel (f) of Fig. \ref{fig:parameters}, lowering $\epsilon_e$ by one order of magnitude would lower the radio flux by about one order of magnitude. Therefore, in this case the allowed parameter space in the $n-\epsilon_B$ plane is greatly enlarged.

\begin{figure}[tbph]
	\begin{center}
		\includegraphics[width=0.87\textwidth,angle=0]{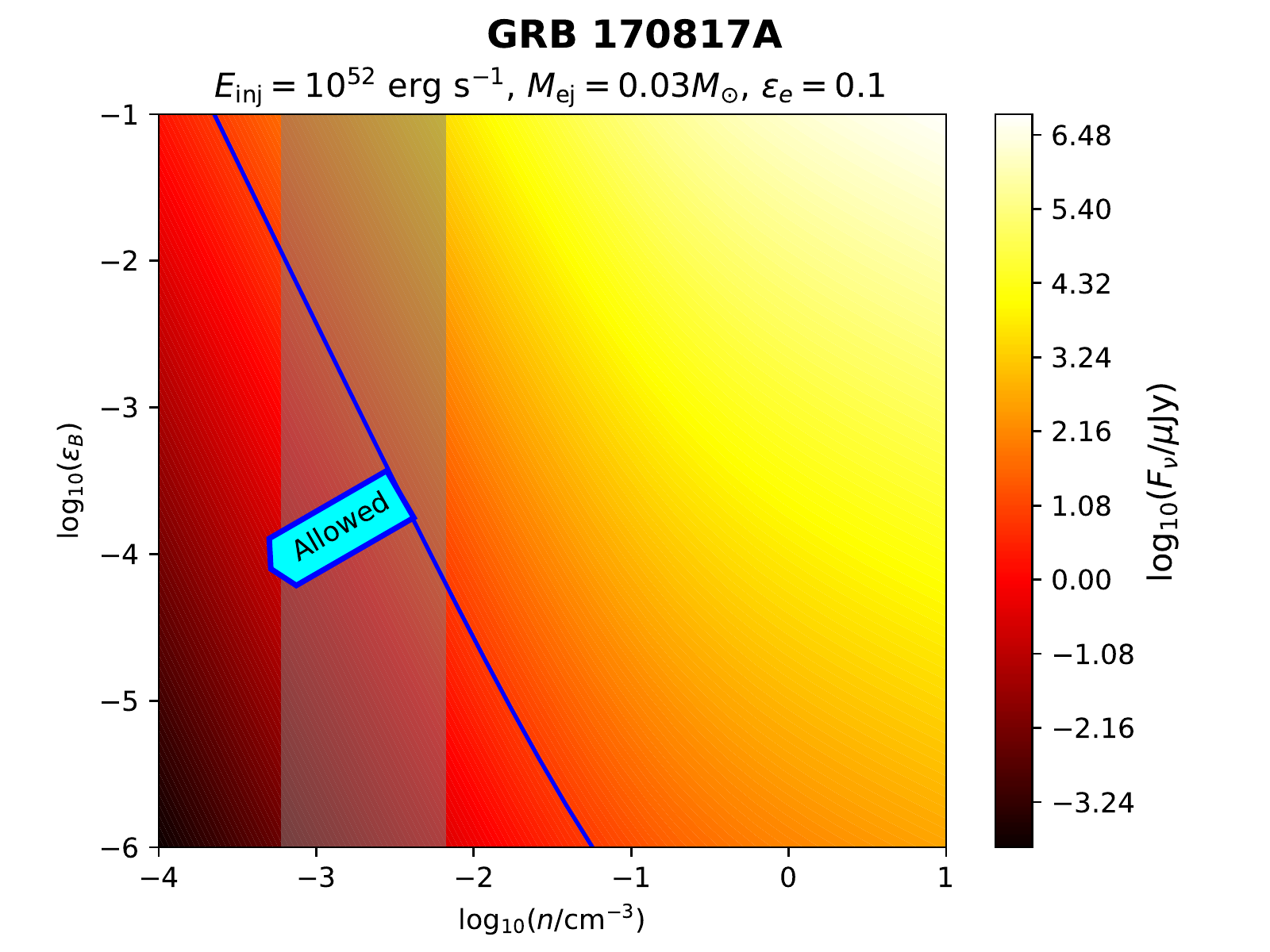} 
	\end{center}
	\caption{Constraint on the parameter space of GRB 170817A.}
	\label{fig:contour-GRB170817A}
\end{figure}

Recently, \cite{Hajela2019} presented the VLA observations of GW 170817 at $\sim$ 2 years after the merger, they obtained  the upper limit flux at 6 GHz as $F_\nu = 8.4 \mu$Jy. By modeling the thermal UV-optical-NIR kilonova (AT 2017gfo) associated with GW 170817, \cite{Villar2017} constrained the total ejecta mass $M_{\rm ej} \sim 0.08M_{\odot}$ within the radioactive-power-dominated scenario. Such a high value of the ejecta mass is higher than the typical dynamical ejecta obtained by numerical-relativity simulations for binary neutron star mergers \citep{Shibata2017}.  The spin-down of the long-lived remnant NS offers additional energy to power the kilonova. Therefore, the required mass of the merger ejecta could be somewhat smaller than that required by the single radioactive power model \citep{Yu2018,Li2018,Ai2018}. By invoking the energy from the long-lived remnant NS, a relatively normal ejecta mass of $M_{\rm ej}=0.03 \pm 0.002M_{\odot}$ could account for the kilonova. \cite{Hajela2019} modeled the broadband afterglow of GRB 170817A, indicating the circumburst medium density $n = 2.5^{+4.1}_{-1.9} \times 10^{-3}$ cm$^{-3}$.  In Fig. \ref{fig:contour-GRB170817A}, we show the constraint on the parameter space for GRB 170817A, indicating that there is still a reasonably large parameter space to allow the existence of a long-lived NS with $E_{\rm rot} \sim 10^{52} \ {\rm erg}$ to satisfy the radio observation constraint. 

\begin{figure*}\label{fig:sample}
	\begin{minipage}[c]{\textwidth}
		\tabcolsep0.0in
		\includegraphics*[width=0.31\textwidth,clip=]{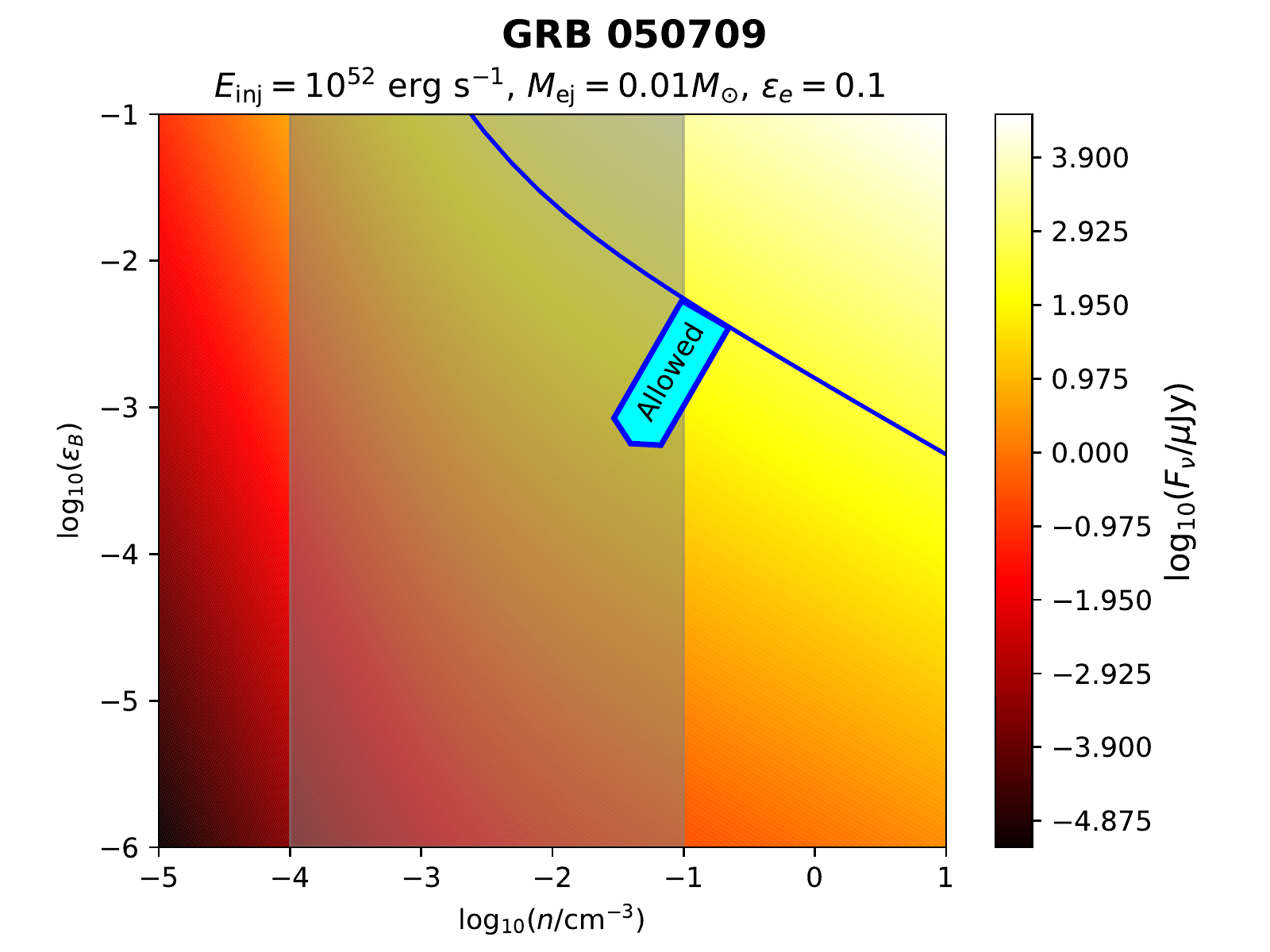}
		\includegraphics*[width=0.31\textwidth,clip=]{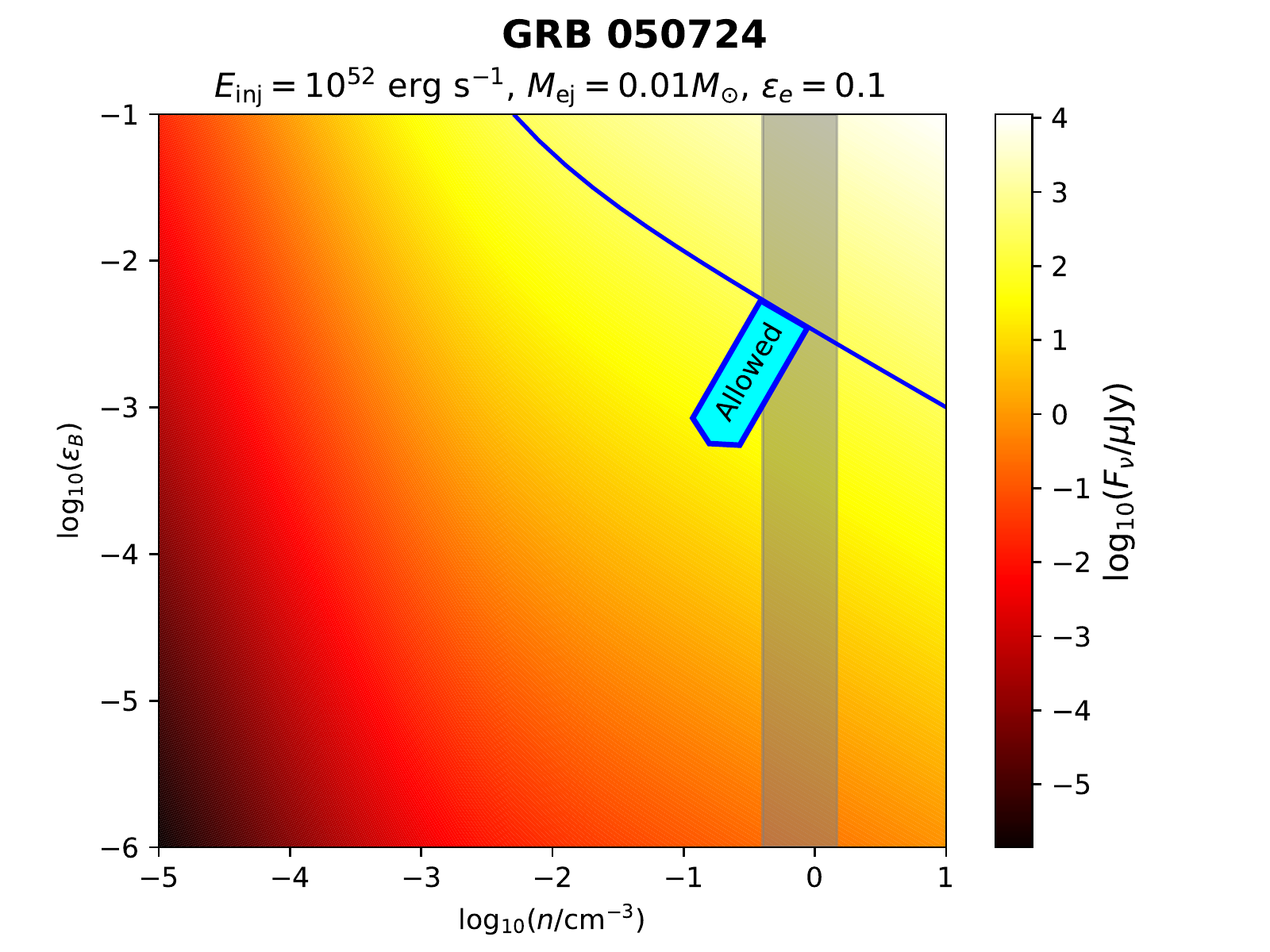}
		\includegraphics*[width=0.31\textwidth,clip=]{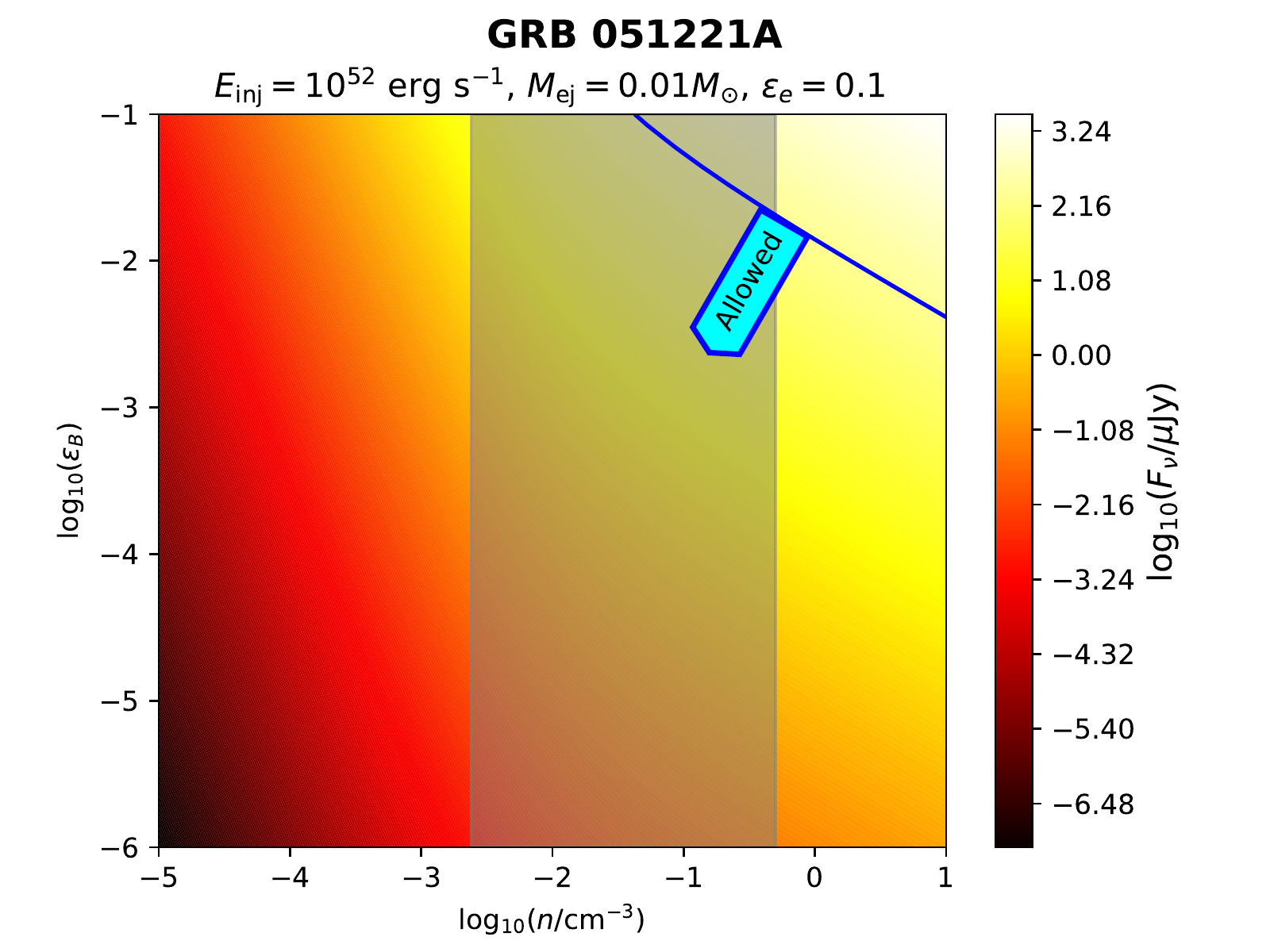} \\
		\includegraphics*[width=0.31\textwidth,clip=]{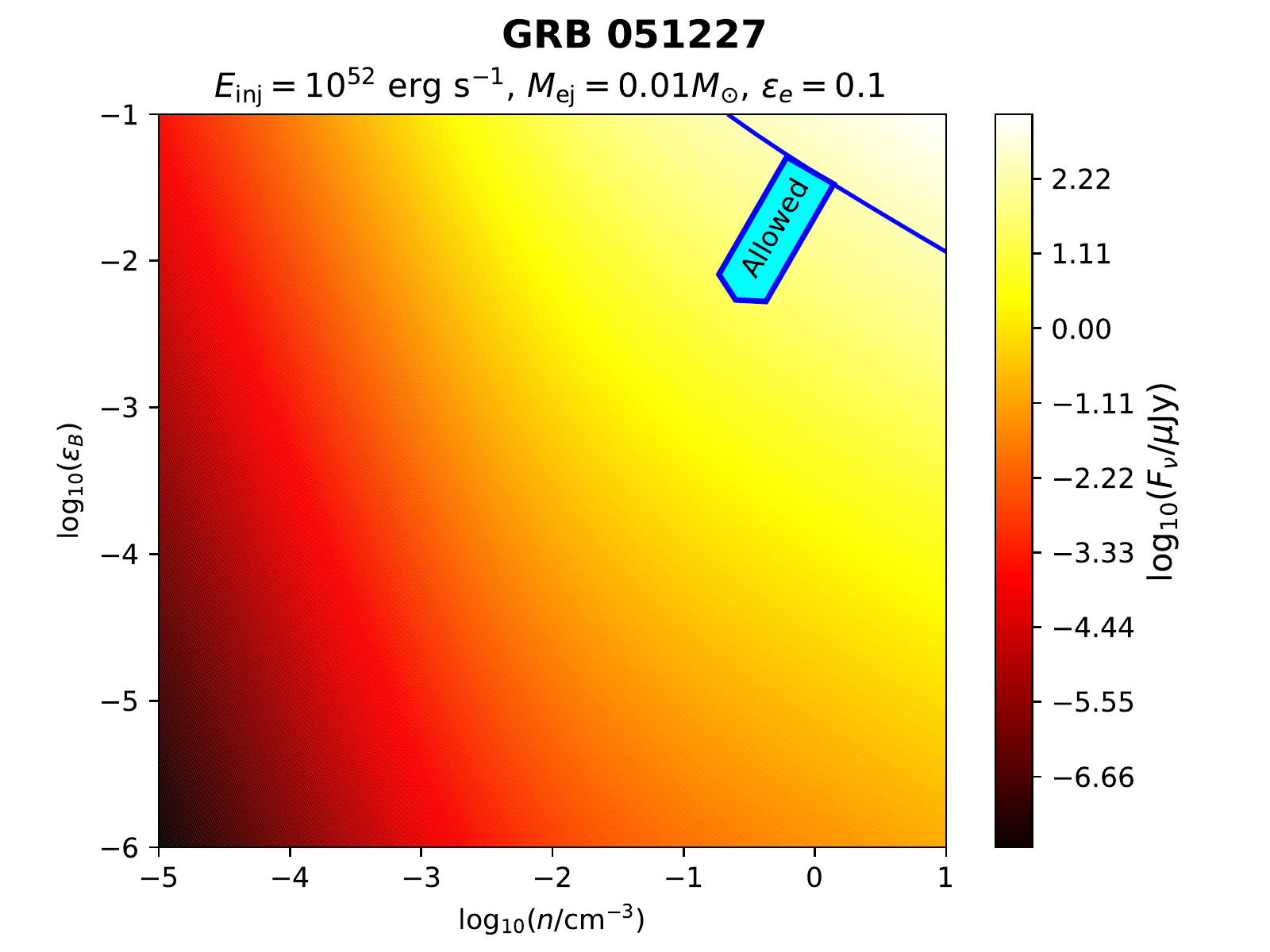}
		\includegraphics*[width=0.31\textwidth,clip=]{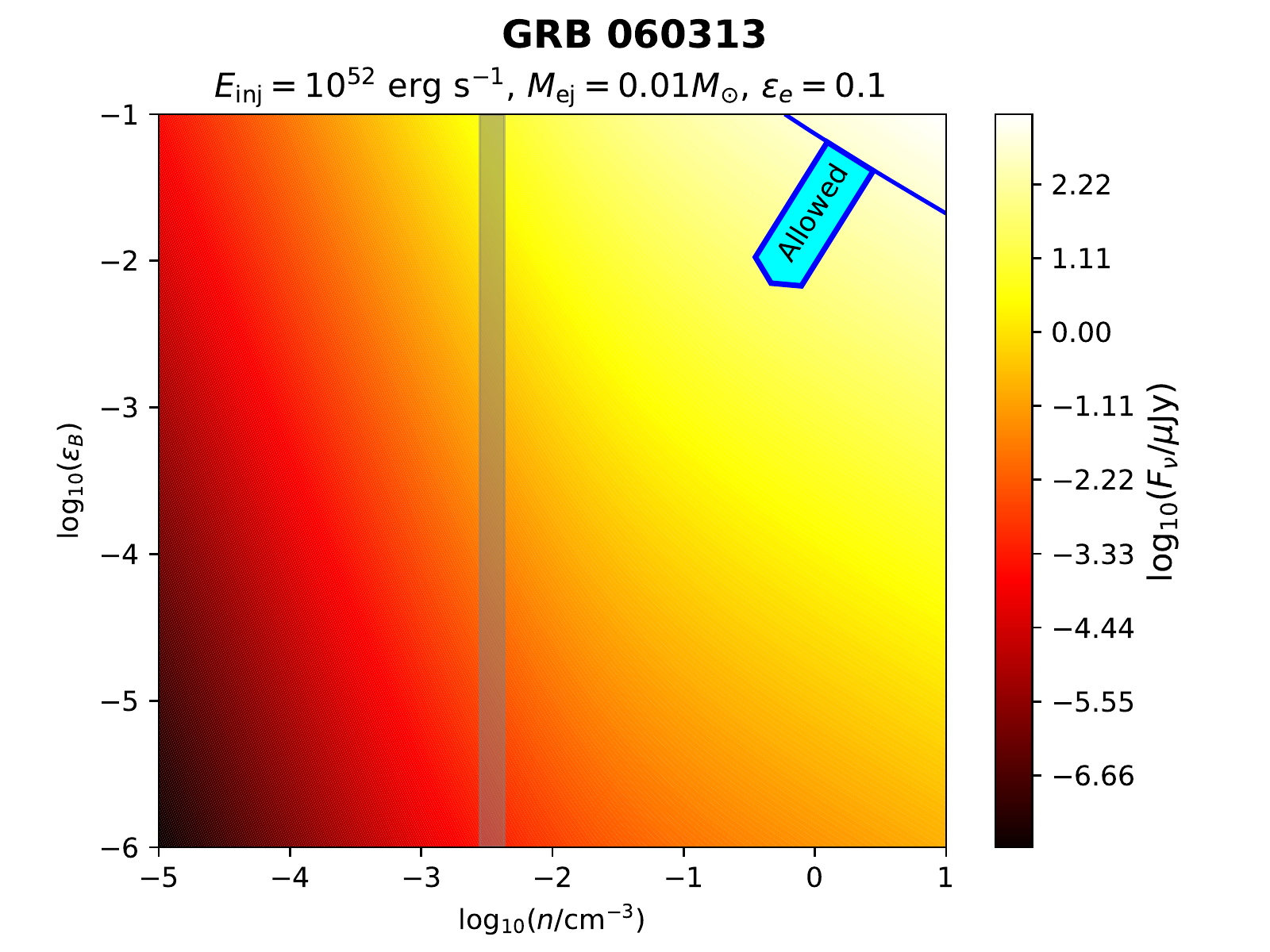}
		\includegraphics*[width=0.31\textwidth,clip=]{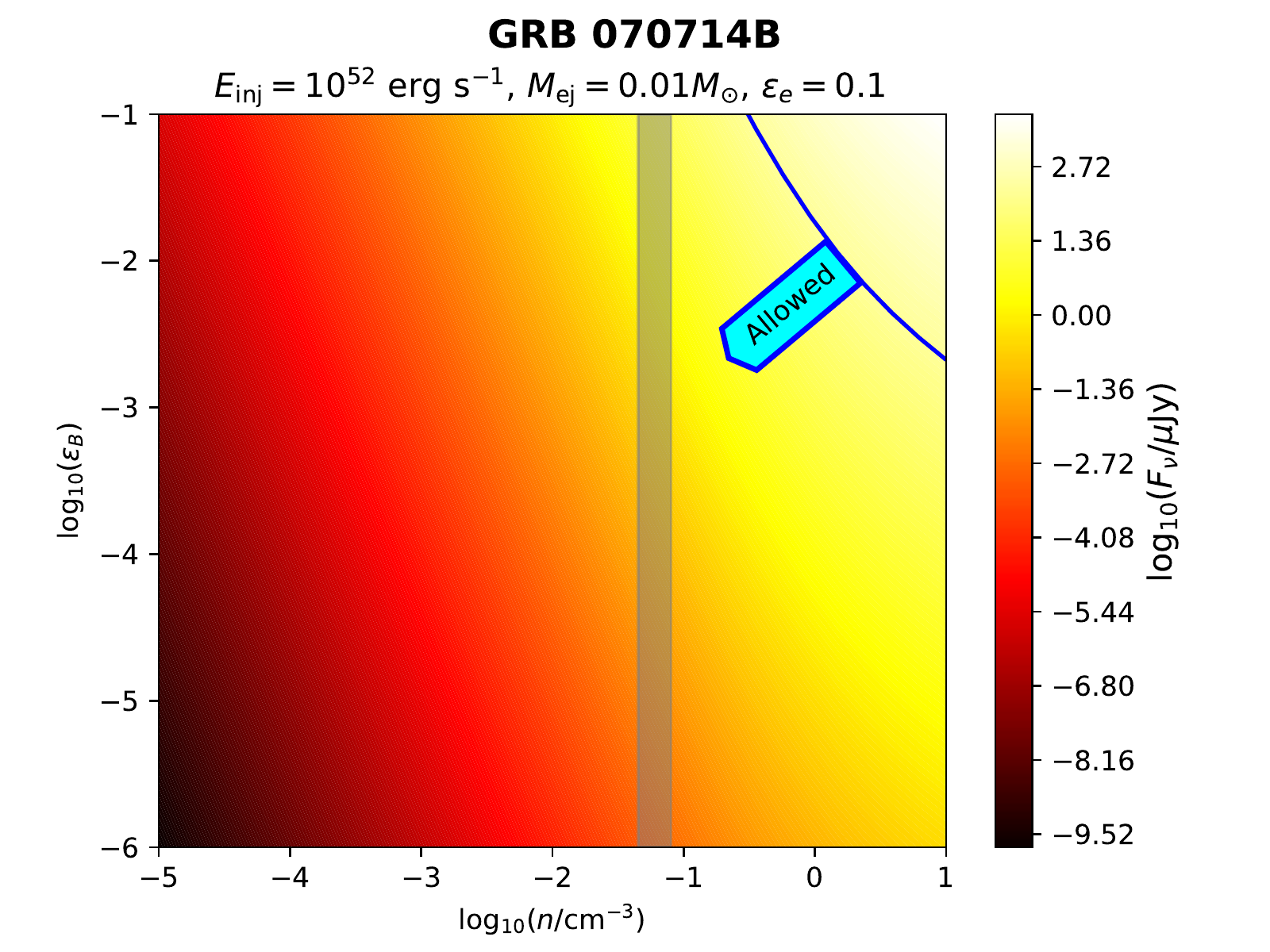}\\
		\includegraphics*[width=0.31\textwidth,clip=]{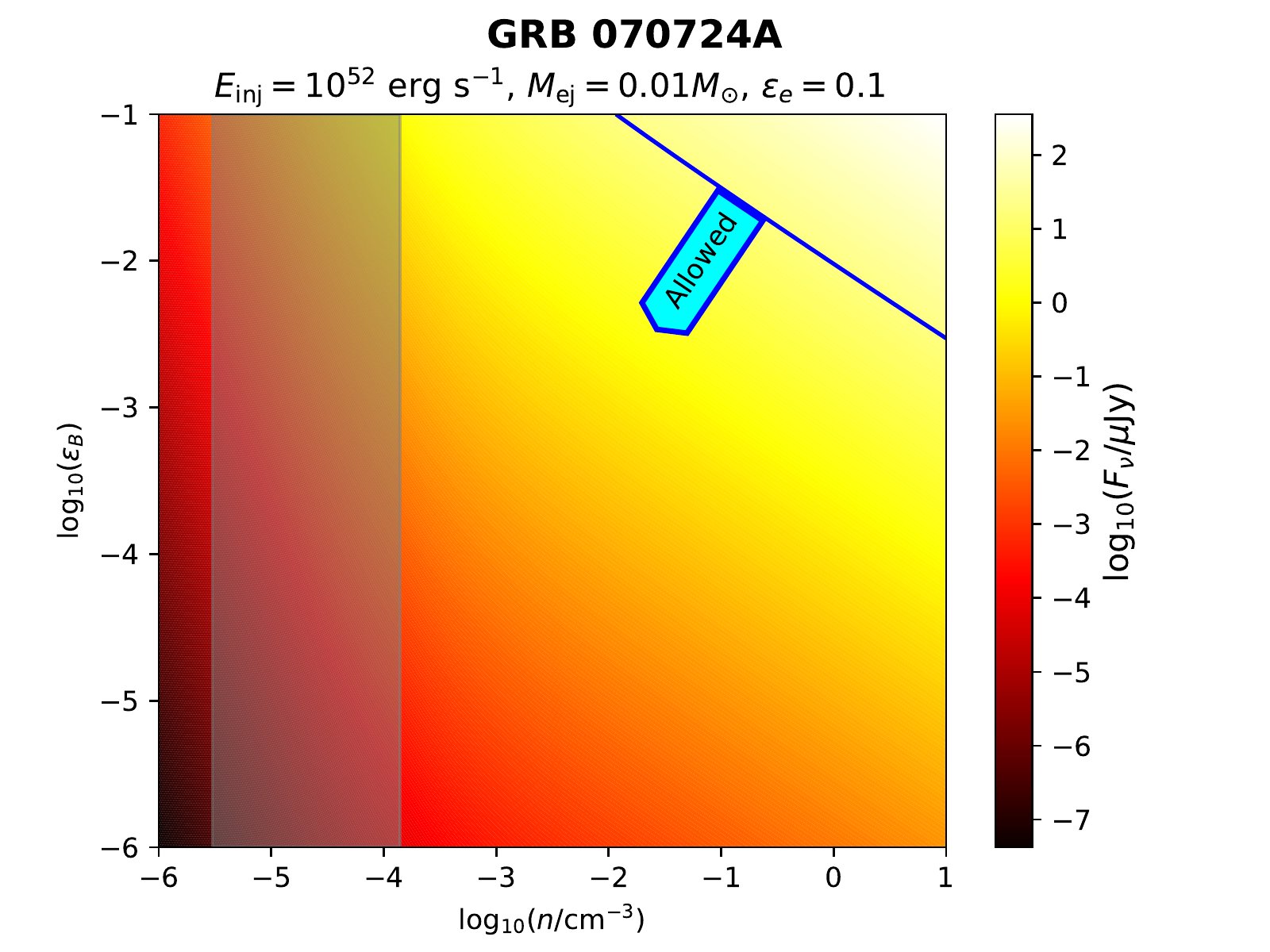}
		\includegraphics*[width=0.31\textwidth,clip=]{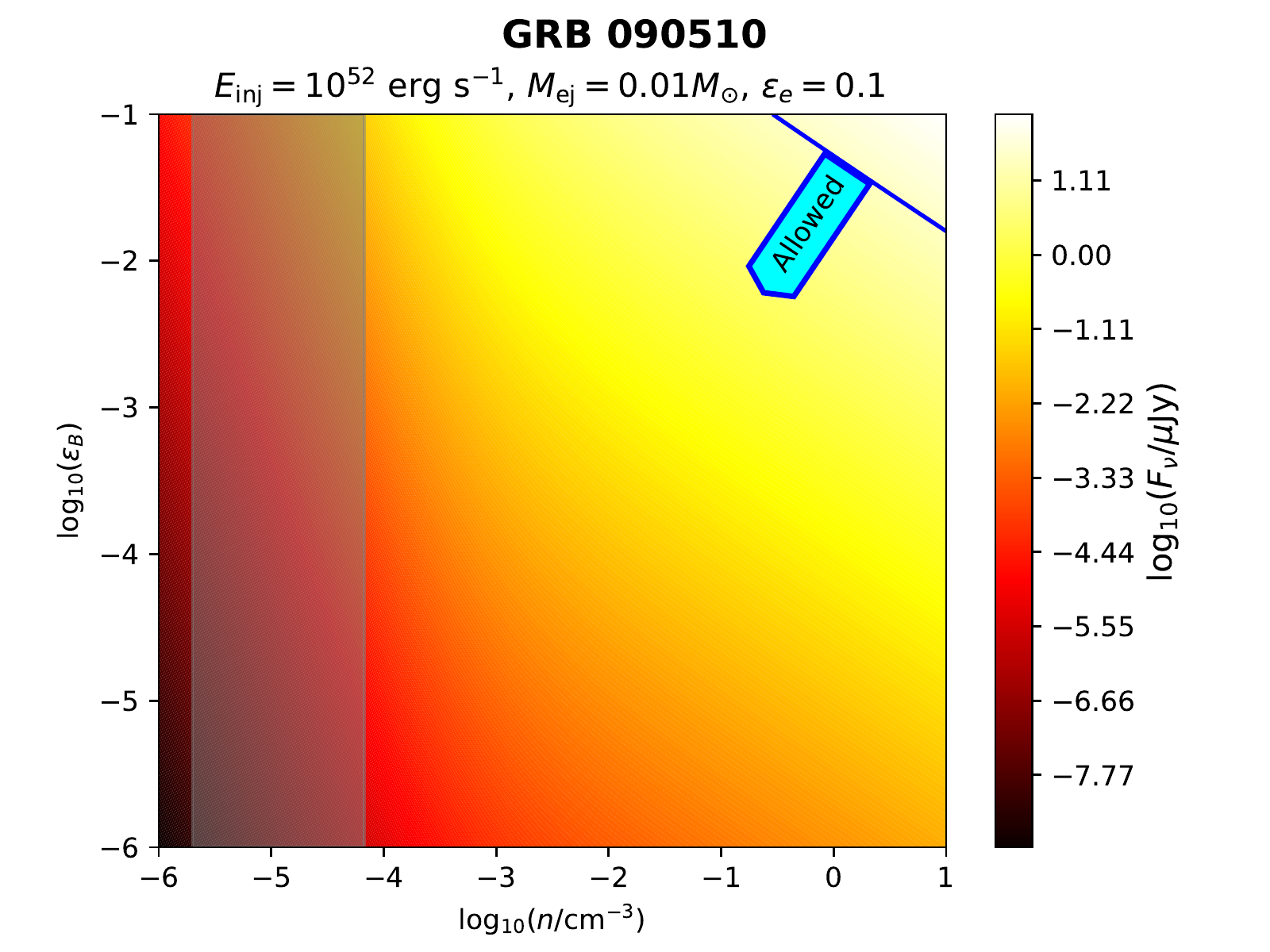}
		\includegraphics*[width=0.31\textwidth,clip=]{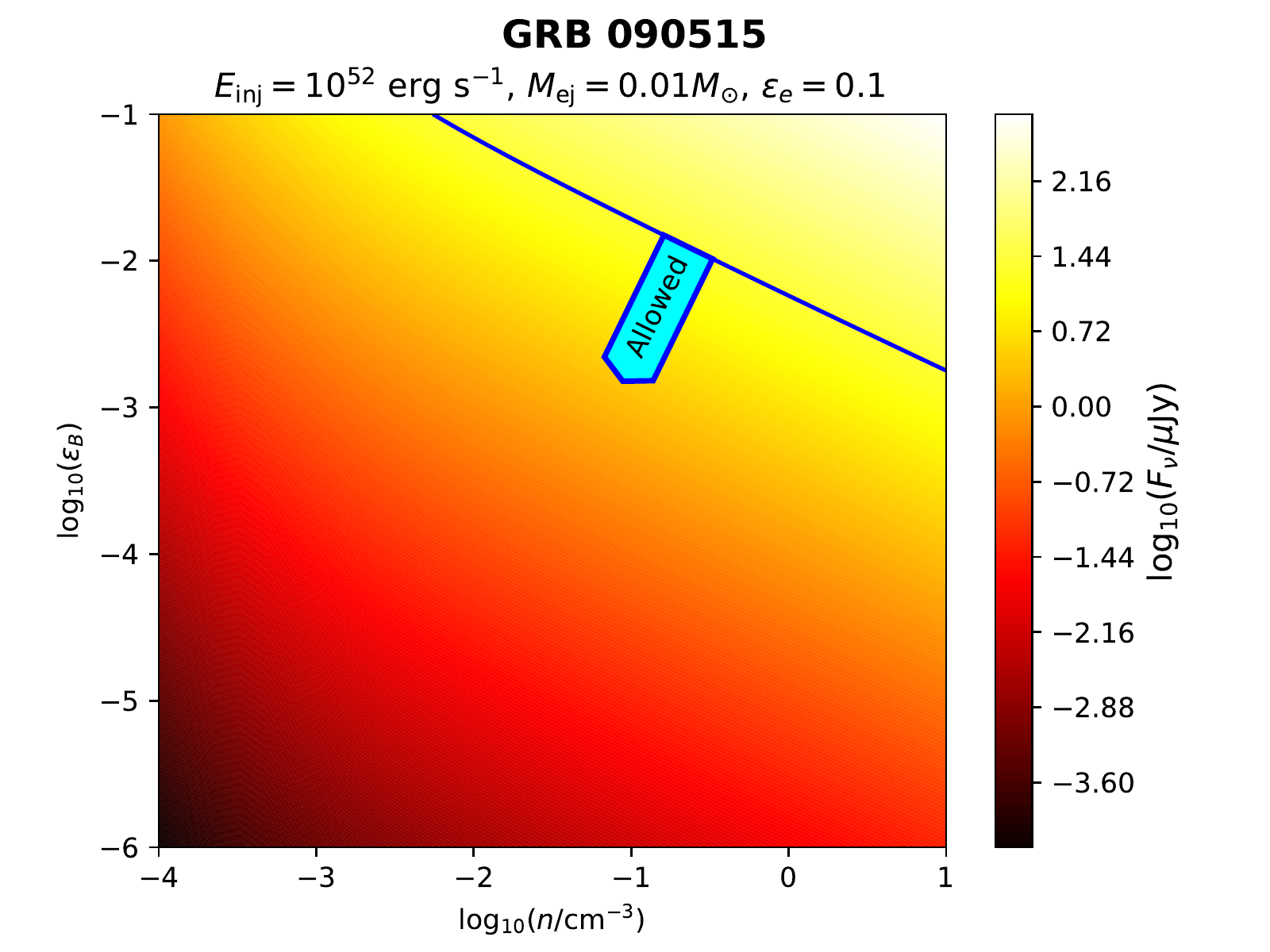}\\
		\includegraphics*[width=0.31\textwidth,clip=]{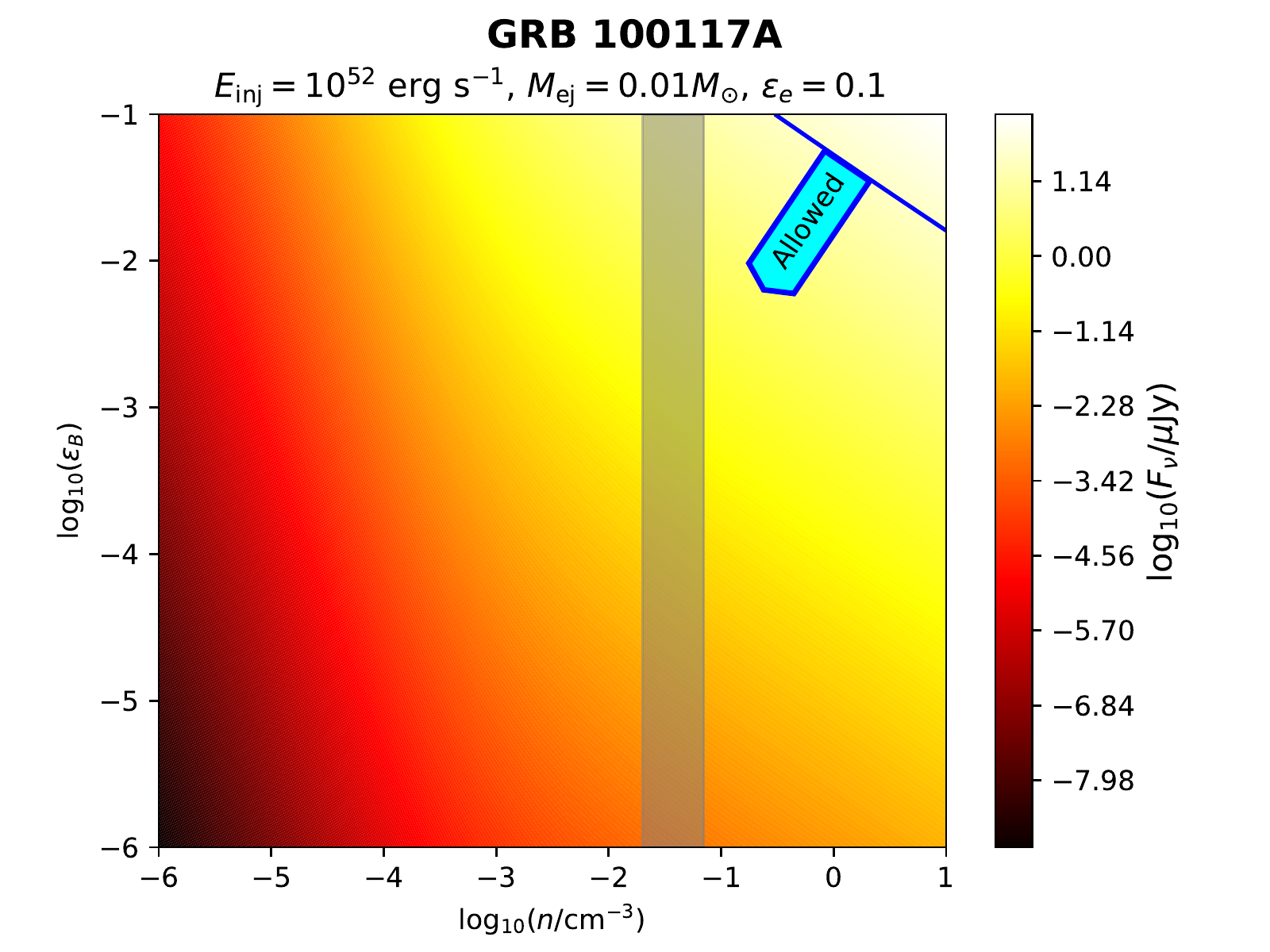}
		\includegraphics*[width=0.31\textwidth,clip=]{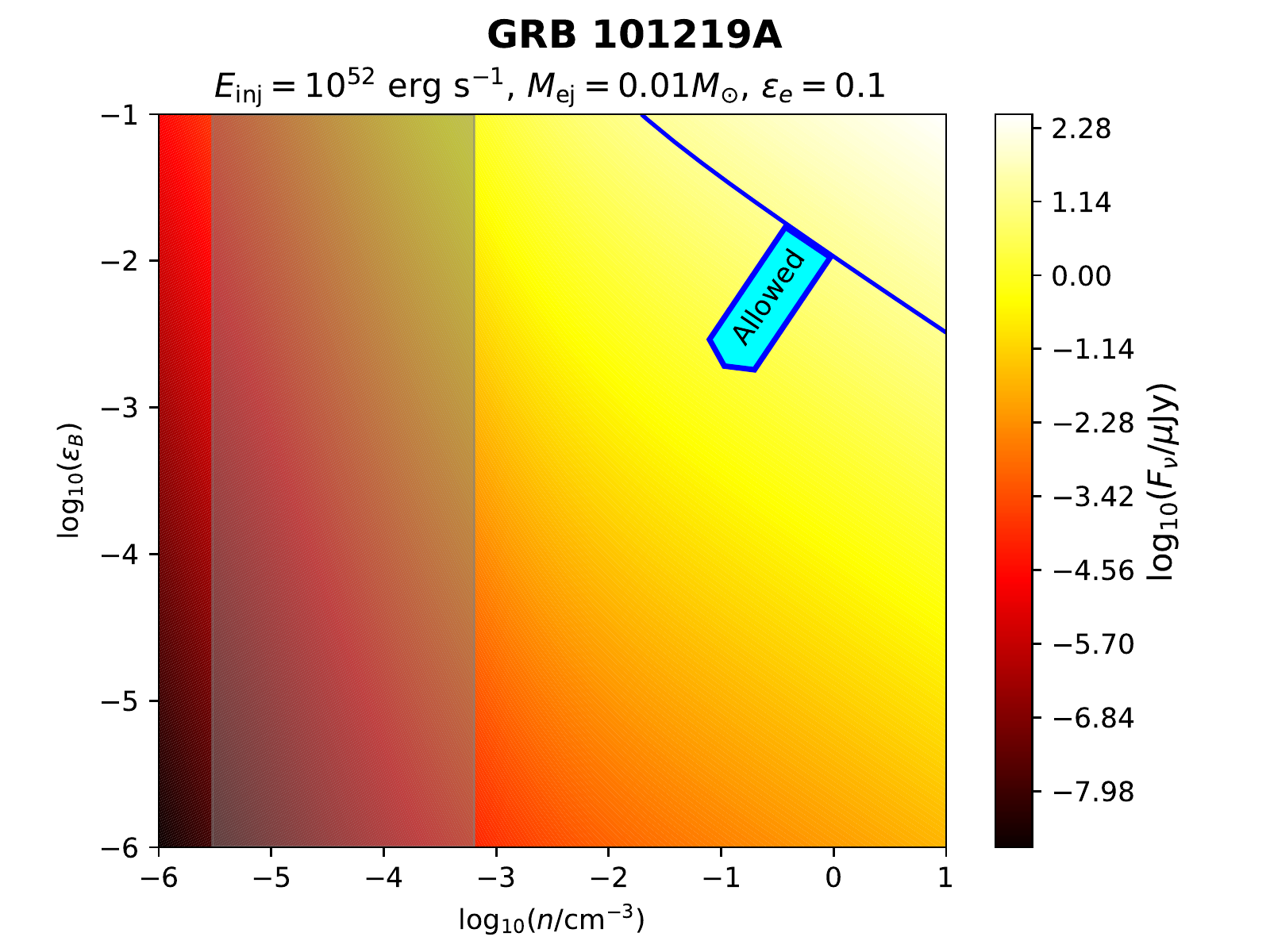} 
		\includegraphics*[width=0.31\textwidth,clip=]{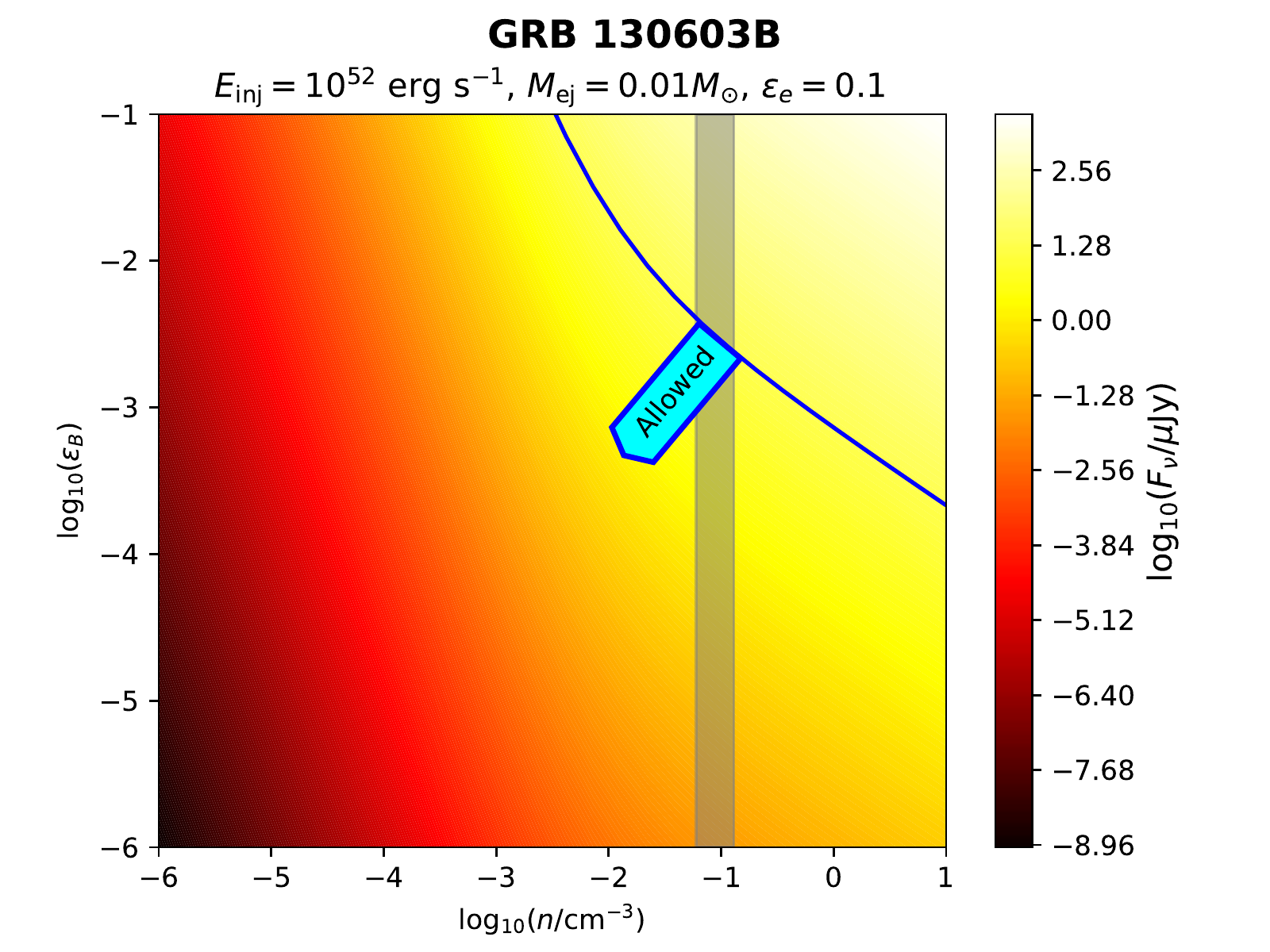}

	\end{minipage}
	\caption{Constraints on the parameter space for the rest 12 SGRBs in our sample.  The parameters are fixed as the following values: $E_{\rm inj} =10^{52}$ erg,  $M_{\rm ej}=10^{-2}M_{\odot}$, $L_{\rm sd,0}=10^{48}$ erg s$^{-1}$, $\epsilon_e = 0.1$, and $p=2.3$. }
\end{figure*}

Broadband modeling of the SGRB afterglows could provide the measurements of the circumburst density and the shock microphysics parameters, which can be used as independent constraints on the allowed parameter space for the magnetar model.
SGRBs prefer to occur in relatively low density environments with a median surrounding circumburst density of $n \sim 4 \times 10^{-3}$ cm$^{-3} $ \citep{Fong2015}. There is a  narrow distribution of the $\epsilon_e$ values from the literature. About $62\%$ of the GRBs in the sample adopted by \cite{Santana2014} have $\epsilon_e \sim 0.1-0.3$. It seems likely that $\epsilon_e$ does not change by much from burst to burst. However, there is a much wider range in the distribution of  $\epsilon_B$ values. \cite{Santana2014} did a systematic study on the magnetic fields in GRB external shock based on a large X-ray and optical afterglow sample and found that the distribution of  $\epsilon_B$ has a range of $\sim 10^{-8}-10^{-3}$ with a median value  of $\sim$ few $\times 10^{-5}$.  \cite{Gao2015} found that the value of the magnetic equipartition parameter in the external shock ranges from $10^{-6}$ and $10^{-2}$.
We collect the inferred densities $n$ for each burst from the literature, the values are listed in Table \ref{tb:epsilonB}. The constraints on the parameter space for the rest 12 SGRBs in our sample are shown in Fig. \ref{fig:sample}. There is no detailed modeling of the afterglows of GRB 0051227 and GRB 090515 to constrain the densities. Assuming that a magnetar injects $10^{52}$ erg of the rotational energy into the surrounding medium and an ejecta mass $M_{\rm ej}=0.01M_{\odot}$, the maximum allowed values of $\epsilon_{B}$ are listed in Table \ref{tb:epsilonB}. We find that  the constraints on the maximum $\epsilon_{B}$ for GRB 060313, 070714B, 070724A, 090510, and 101219A reach the upper limit we set a prior. For all 15 SGRBs with non-detection of the radio emission, the constraints on the upper limit of $\epsilon_{B}$ in the magnetar scenario are consistent with the expectations from the modeling of GRB afterglows.


%
%
%
%
%

\section{Conclusions and Discussion}
\label{Sec: Discussion}

A long-lived magnetar remnant has been wildly invoked to explain the observational properties of the
X-ray afterglows of SGRBs. Late-time radio observations of SGRBs provide a potential way to place a constraint on the existence of a long-lived magnetar remnant.  We developed a sophisticated model to calculate the radio emission from the interaction between the merger ejecta and the circumburst medium in the magnetar scenario. Our model invokes several important physical processes, e.g, generic hydrodynamics, relativistic effects, and the deep Newtonian phase.  The theoretical light curves predicted by our model in the timescale of $\sim 1-10$ yr is about one order of magnitude lower than those predicted in previous oversimplified models \citep{Metzger2014,Fong2016}, which used the non-relativistic calculations following \cite{Nakar2011} but with a higher kinetic energy ($\sim 10^{52}$ erg) of the ejecta. 
Our generic dynamical model applies to both the relativistic and the non-relativistic phases. Our calculations also extend to the deep Newtonian phase
when the minimum Lorentz factor of the electrons, $\gamma_m$, drops below to unity
which results in a shallower decline rate of the light curve.   

 We collected 15 SGRBs late-time radio observational data from the literature \citep{Metzger2014,Horesh2016,Fong2016,Hajela2019}. All the events show an X-ray emission signature (e.g. X-ray plateau, extended emission or X-ray flares) that may be interpreted as being powered by a magnetar central engine. No radio source was detected from any GRB in our sample. We derive the constraints on the maximally allowed $\epsilon_{B}$ in the magnetar scenario.
 Our results show that all the non-detections can be accommodated within the magnetar engine model with a reasonably large allowed parameter space, which also overlaps with that inferred from SGRB afterglow modeling. Considering the possibility of low values of shock microphysics parameters as inferred from GRB multi-band afterglow observations and the simplified modeling by previous authors, the radio upper limits reported in previous works \citep{Metzger2014,Fong2016,Horesh2016}  may not necessarily pose severe constraints on the existence of a long-lived magnetar remnant in these short GRBs.
 
More extreme parameters (e.g. $E_{\rm rot} \sim 10^{53}$ erg for a $\sim 2.2M_{\odot}$ NS with a spin period close to $1$ ms) are ruled out for some bursts. However, it is unlikely that a new-born supramassive NS can ejecta a kinetic energy of such an order. The newborn NS may possess a large ellipticity, which would release energy through secular gravitational waves  \citep{Gao2016,Ai2018}. Strong GW emission is expected in the post-merger phase, due to deformations of the core caused by  the high magnetization \citep[e.g.][]{Dall2015}. A long-lived remnant with a typical injected energy $E_{\rm rot} \sim 10^{52}$ erg may be more likely. Such magnetars are generally allowed for all the 15 SGRBs studied in our sample.

Future radio  telescopes such as SKA and ng-VLA with their sub-$\mu$Jy level sensitivity will be able to improve the current limits of the afterglows. The detection of late-time radio emission from the interaction the merger ejecta with the circumburst  medium would confirm the existence of a long-lived magnetar remnant. Non-detections, on the other hand, would substantially tighten the parameter space allowed by the magnetar model, and rule out the existence of such an engine in some cases.

%
%
%
%
%

\acknowledgments
We thank the referee  for his/her valuable suggestions.
We thank Yun-Wei Yu for helpful discussions. This work is supported by the National Natural Science Foundation of China (NSFC) under Grant No. 11722324,11690024,11603003,11633001, the Strategic Priority Research Program of the Chinese Academy of Sciences, Grant No. XDB23040100 and the Fundamental Research Funds for the Central Universities. LDL is supported by  the National Postdoctoral Program for Innovative Talents (grant No. BX20190044), China Postdoctoral Science Foundation (grant No. 2019M660515)  and ``LiYun'' postdoctoral fellow of Beijing Normal University.

\clearpage

%

\begin{table}[tbph] \label{tb:models}
	\caption{Comparisons of our model with four previous works.}
	\begin{tabular}{lccccc}
		\hline  \hline
		& This paper & Horesh2016 & Fong2016 & Metzger2014 & Nakar2011\\ \hline
		Energy injection from magnetar     & $\checkmark$ & $\checkmark$ & $\checkmark$ & $\checkmark$ & $\times$\\
		Synchrotron self-absorption        &$\checkmark$  &$\checkmark$  & $\checkmark$ & $\checkmark$ & $\checkmark$ \\
		Generic hydrodynamic               & $\checkmark$ & $\checkmark$ & $\times$     & $\times$     & $\times$\\
		Doppler effect                      & $\checkmark$ & $\checkmark$ & $\times$     & $\times$     & $\times$\\
		Deep Newtonian phase               & $\checkmark$ & $\times$     & $\times$     & $\times$     & $\times$\\
		\hline
	\end{tabular}
\par
The ``$\checkmark$'' sign denotes that this physical process has been invoked in the corresponding model, and the ``$\times$'' sign represents that this process was ignored.
\end{table}

\clearpage

\begin{table}[tbph] \label{tb:sample}

		\caption{Late-time radio observations of SGRBs in our sample} 
		\begin{center}
	\begin{tabular}{lccccccc}
		\hline \hline
		GRB & $z$ & $\nu_{\rm{obs}}$ & $T_{\rm{obs}} $ \tablenotemark{a} & $F_{\nu} $ \tablenotemark{b}  & $\nu
		L_{\nu}$  & X-ray behavior & Reference\tablenotemark{c} \\
		\  & \  & (GHz) & (days) &  ($\mu$Jy) & ($10^{38}$ erg s$^{- 1}$) & \  & \
		\\
		\hline
		050709 & 0.16 & 1.4 & 924 & 350 & 3.4 & Extended emission & 1\\
		
	    \multirow{2}*{050724\tablenotemark{d}}  & \multirow{2}*{0.257} & 1.4 & 913 & 240 & 6.7 &  \multirow{2}*{ Extended emission} &   \multirow{2}*{1,2}\\
		~                     & ~ & 6.0 & 3500 & $22.1$ & $2.7$ & ~ & ~\\
		
		 \multirow{2}*{051221A\tablenotemark{d}}  &  \multirow{2}*{0.547} & 1.4 & 759 & 210 & 34.7 &  \multirow{2}*{Extended emission/Plateau} &  \multirow{2}*{1,2}\\
		~ & ~ & 6.0 & 3350 & 19.5 & 14 & ~ & ~\\
		
		051227 & 0.8 & 1.4 & 753 & 240 & 101.5 & Extended emission & 1\\
		060313 & 0.75  & 1.4 & 677 & 510 & $183.4$ & Extended emission & 1\\
		060505 & 0.089 & 1.4 & 624 & 330 & 0.89 & Extended emission & 1\\
		070714B & 0.923 & 1.4 & 189 & 190 & $114.7$ & Extended emission & 1\\
		070724A & 0.457 & 6.0 & 2768 & 19.1 & $9.1$ & Plateau & 2\\
		080905A & 0.122 & 6.0 & 2363 & 22.2 & $0.52$ & Plateau & 2\\
		090510 & 0.903 & 6.0 & 2127 & 26.5 & $66$ & Extended emission & 2\\
		090515 & 0.403 & 6.0 & 2117 & 22.7 & $8.0$ & Plateau & 2\\
		100117A & 0.915 & 6.0 & 1867 & 32 & $83$ &  Plateau \& Flares & 2\\
		101219A & 0.718 & 6.0 & 1528 & 17.5 & $2$5 &  Plateau & 2\\
		 \multirow{2}*{130603B$^{\rm d, e}$}  &  \multirow{2}*{0.356} & 3.0 & 619 & 60 & 8.6 &  \multirow{2}*{Excess emission/Plateau} &  \multirow{2}*{2,3}\\
		~ & ~ & 6.0 & 639 & 20.6 & $5.4$ & ~ &~\\
		170817A\tablenotemark{f} & 0.00978 & 6.0 & 724 & 8.4 & $1.18 \times 10^{-3}$ &  Extended emission & 4\\
		
		\hline
	\end{tabular}
	\end{center}	
	a. $T_{\rm{obs}} $ is the observational time after the GRB in observer frame, the rest frame time after burst $T_{\rm rest}=T_{\rm obs}/(1+z)$.
	
	b. The upper limit flux $F_{\nu}$ inferred by non-detection of late-time radio emission.
	
	
	c. References  for radio observations: (1) \cite{Metzger2014}, (2) \cite{Fong2016}, (3) \cite{Horesh2016}, and (4) \cite{Hajela2019}.
	
	d. GRB 050724, GRB 051221A, and GRB 130603B have twice radio observations on different frequencies and different times.
	
	e. GRB 130603B is a possible kilonova candidates.
	
	f. \cite{Piro2019} reported a low-significance X-ray variability in GRB 170817A at 155 days after the merger.
	
\end{table}

\clearpage
\begin{table}[tbph]
	\label{tb:epsilonB}
	\caption{Constraints on the magnetic equipartition parameter  $\epsilon_B$.}
	\begin{tabular}{lccl}
		\hline \hline
		GRB & $n$ \tablenotemark{a} & $\epsilon_{B,\max}$  \tablenotemark{b} & Reference \tablenotemark{c} \\
		\  & (cm$^{-3}$) & & \\
		\hline
		050709 & $10^{-4}-0.1$ & $3.6 \times 10^{-2}$ & \cite{Panaitescu2006} \\
		050724 & $0.4-1.47$  & $5.7 \times 10^{-2}$ & \cite{Fong2015} \\
		051221A & $2.4 \times 10^{-3}-0.5$  & $1.5 \times 10^{-3}$ & \cite{Soderberg2006}\\
		051227 \tablenotemark{d} &           $-$ & $-$  &  $-$\\
		060313 \tablenotemark{e} & $3.3^{+1.0}_{-0.5}\times10^{-3}$   & 0.1 & \cite{Fong2015}\\
		060505 & $1.0$  & $3.7\times 10^{-4}$ & \cite{Xu2009} \\
		070714B \tablenotemark{e}& $5.6^{+2.4}_{-1.1}\times10^{-2}$   &  0.1 &\cite{Fong2015}\\
		070724A & $1.9^{+12}_{-1.6}\times 10^{-5}$   & 0.1  &\cite{Fong2015}\\
		080905A &  $1.3^{+33}_{-1.2} \times10^{-4}$  & $7.6 \times 10^{-3}$ & \cite{Fong2015}\\
		090510 \tablenotemark{e} & $1.2^{+5.5}_{-1.0} \times 10^{-5}$   & 0.1 &\cite{Fong2015}\\
		090515 \tablenotemark{d} & $-$   &$-$ &$-$\\
		100117A \tablenotemark{e}  & $4.0^{+3.0}_{-1.0}\times 10^{-2}$  & 0.1  &\cite{Fong2015}\\
		101219A \tablenotemark{e}  & $4.6^{+59}_{-4.3}\times 10^{-5}$  & 0.1 &\cite{Fong2015}\\
		130603B & $9.0^{+4.0}_{-3.0}\times 10^{-2}$   & $2.3 \times 10^{-3}$ &\cite{Fong2015} \\
		170817A \tablenotemark{f} & $2.5^{+4.1}_{-1.9} \times 10^{-3}$   & $8.6 \times 10^{-4}$ & \cite{Hajela2019} \\
		
		 \hline
	\end{tabular}
\\	

a. The circumburst density $n$ based on GRB afterglow modeling from the literature.

b. Maximum  allowed $\epsilon_B$ by the observation assuming $E_{\rm inj} =10^{52}$ erg,  $M_{\rm ej}=10^{-2}M_{\odot}$, $L_{\rm sd,0}=10^{48}$ erg s$^{-1}$, $\epsilon_e = 0.1$, and $p=2.3$, and we adopt the maximum value of $n$.

c. References for the circumburst densities.

d. No afterglow modeling were available for GRB 051227 and GRB 090515.

e. The constraints on maximum allowed $\epsilon_{B}$ for GRB 060313, 070714B, 070724A, 090510, and 101219A reach the upper limit we set prior.

f. For GRB 170817A, we adopt $M_{\rm ej} = 0.03M_{\odot}$, while the rest of the parameters are kept fixed.
\end{table}

\clearpage

\clearpage

\clearpage

\clearpage

\clearpage

\clearpage

\end{document}